\documentclass[
    aps,
    pra,
    reprint,
    superscriptaddress,
    amsmath,
    amssymb,
    longbibliography
]{revtex4-2}

\usepackage{graphicx}   
\usepackage{dcolumn}    
\usepackage{bm}         
\usepackage{xcolor}     
\usepackage{hyperref}   
\usepackage{eucal}
\usepackage{braket}
\newcommand{\A}{\mathcal{A}}
\newcommand{\B}{\mathcal{B}}
\newcommand{\C}{\mathcal{C}}
\newcommand{\bilin}[1]{\langle\!\langle #1 \rangle\!\rangle}
\newcommand{\eps}{\epsilon}

\colorlet{RED}{red}\colorlet{BLUE}{blue}

\begin{document}

\title{Biorthogonality of Guided Modes in Non-Hermitian Electromagnetism}



\author{Amgad Abdrabou}
\affiliation{Department of Electrical and Computer Engineering, Saint Louis University, Saint Louis, MO 63103, USA}

\author{M. A. Swillam}
\affiliation{Department of Physics, The American University in Cairo, New Cairo, 11835, Egypt.}

\author{\c{S}ahin K. \"Ozdemir}
\affiliation{Department of Electrical and Computer Engineering, Saint Louis University,  Saint Louis, MO 63103, USA}

\author{R. El-Ganainy}
\email{relganainy@slu.edu}
\affiliation{Department of Electrical and Computer Engineering, Saint Louis University,  Saint Louis, MO 63103, USA}

\date{\today}

\begin{abstract}
Biorthogonality plays a central role in the analysis of non-Hermitian photonic systems, where modal expansions generally require both right and left eigenstates. While these concepts are straightforward within scalar and coupled-mode descriptions, their interpretation in the full vectorial Maxwell framework is less transparent. In particular, the standard electromagnetic modal orthogonality relation derived from unconjugated Lorentz reciprocity involves both electric and magnetic fields and differs in form from the biorthogonality relations obtained from conventional adjoint-operator arguments. Here, we establish the connection between these two formulations for non-Hermitian waveguiding systems with gain and loss. Starting from the full vectorial electric-field wave equation, we formulate the guided-mode problem as a quadratic eigenvalue problem in the propagation constant, construct its adjoint, and identify the corresponding left eigenmodes. We show that the left eigenmode is related to the complex conjugate of the backward-propagating electric-field mode and demonstrate that the resulting electric-field biorthogonal pairing is equivalent to the mixed electric--magnetic field relation obtained from unconjugated Lorentz reciprocity. The framework also clarifies the electromagnetic meaning of modal normalization, its connection to self-orthogonality at exceptional points, and the recovery of conventional conjugated modal orthogonality in the lossless limit. These results provide a unified operator-theoretic interpretation of electromagnetic biorthogonality and Lorentz reciprocity and establish a direct connection between the left/right eigenstate language of non-Hermitian physics and the reciprocity-based modal framework of classical electromagnetism.
\end{abstract}

\maketitle

\section{Introduction}
Non-Hermitian (NH) optics has emerged as a powerful framework for controlling the flow of light and its interaction with matter~\cite{elganainy2018nhpt,feng2017nhphotonics,miri2019eps,ozdemir2019ptep}. From optical isolation~\cite{peng2014wgm,chang2014isolation} and laser engineering~\cite{feng2014singlemode,hodaei2014microring,bandres2018topological} to nonlinear optics~\cite{ramezani2010unidirectional,lumer2013nonlinear} and sensing~\cite{wiersig2014sensing,hodaei2017higherorder}, the deliberate engineering of gain and loss has enabled functionalities that are difficult, or in some cases impossible, to achieve within conventional Hermitian settings. 

Central to the analysis and design of NH optical systems is an understanding of their linear modal structure and response. A broad range of problems---including the calculation of resolvent operators governing scattering from linear NH systems~\cite{leung1994qnm,ching1998qnm,sauvan2013qnm,lalanne2018qnm}, the evaluation of linewidth enhancement in lasers~\cite{petermann1979,siegman1989}, and the development of coupled-mode descriptions of nonlinear NH systems~\cite{elganainy2007ptcmt,ramezani2010unidirectional,lumer2013nonlinear}---requires the decomposition of an arbitrary optical field in terms of the eigenmodes of the underlying linear system. In Hermitian systems, this decomposition follows directly from the orthogonality of the eigenmodes. In NH systems, however, the eigenmodes are generally nonorthogonal, and the corresponding modal expansion must instead be formulated in terms of both right and left eigenstates, which together form a biorthogonal basis~\cite{curtright2007biorthogonal,brody2014biorthogonal}.

The early development of NH optics was strongly influenced by parity-time ($\mathcal{PT}$)-symmetric quantum mechanics~\cite{bender1998real,bender2007making}. Consequently, much of the early work focused on scalar paraxial wave equations and systems described within coupled-mode theory (CMT)~\cite{haus1991cmt, elganainy2007ptcmt,makris2008beam,guo2009observation}. In both cases, the equations governing optical evolution have a mathematical structure analogous to the single-particle Schr\"odinger equation. This close correspondence made it natural to adopt concepts and mathematical conventions from quantum mechanics, including the conventional inner product and the associated definitions of right and left eigenstates~\cite{mostafazadeh2002pseudo,moiseyev2011}.
\begin{figure*}[t]
    \centering
    \includegraphics[width=\textwidth]{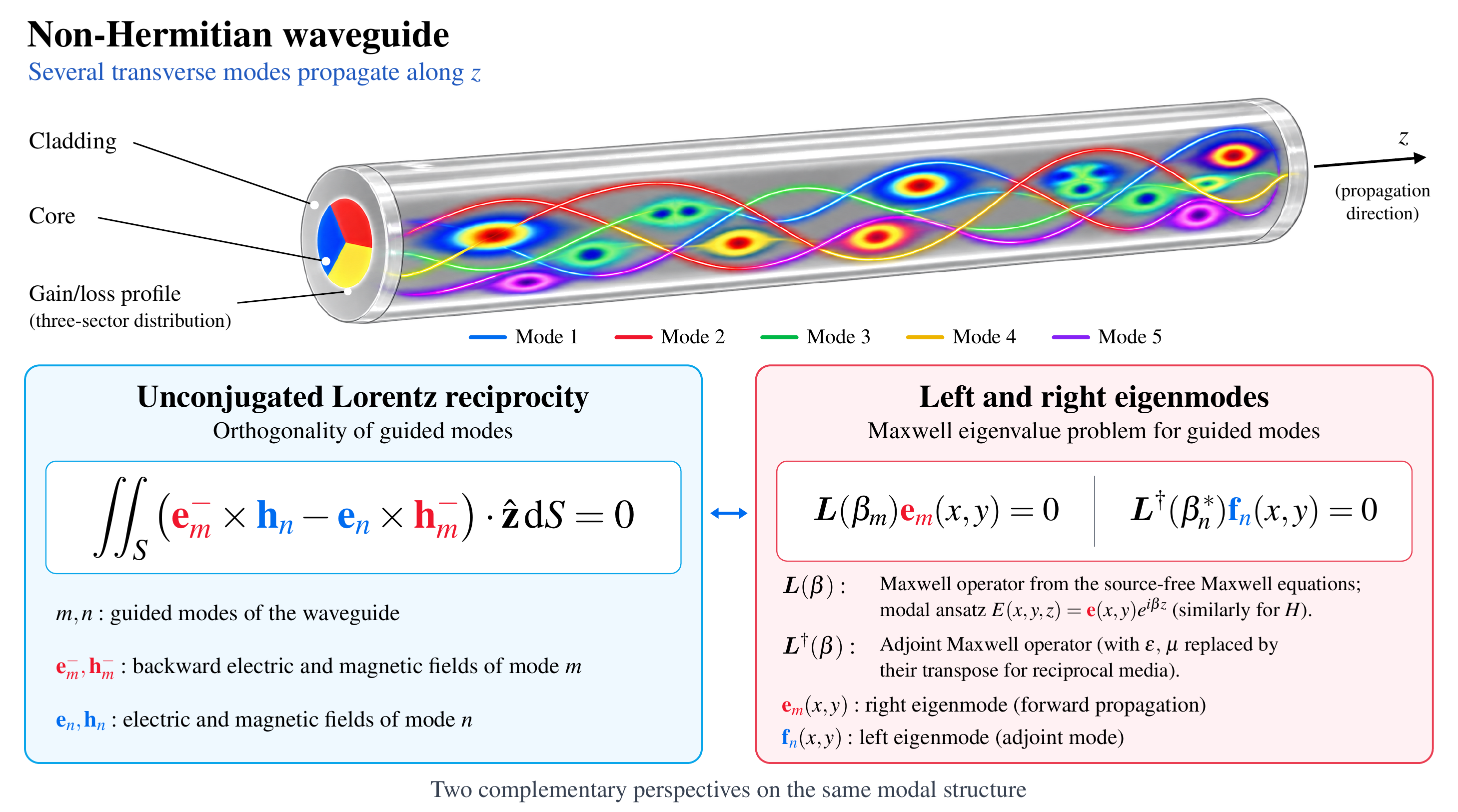}
    \caption{Schematic illustration of a multimode non-Hermitian waveguide with a spatially distributed gain and loss profile. The conventional modal orthogonality relation obtained from unconjugated Lorentz reciprocity involves both the electric and magnetic fields, whereas the operator-theoretic formulation of the Maxwell eigenvalue problem introduces corresponding right and left eigenmodes. The two formulations provide complementary descriptions of the same electromagnetic modal structure.}
    \label{fig:Schematic}
\end{figure*}

However, models based on scalar approximations or CMT remain approximations to the full vectorial Maxwell equations. In many situations, these approximations are not adequate, and a fully vectorial description is required~\cite{afshar2009vectorial,driscoll2009longitudinal,turner2009vectorial}. In this case, the corresponding biorthogonal structure must also be properly established. Since Maxwell's equations can be reduced to a wave equation involving only the electric or magnetic field, it is tempting to proceed in direct analogy with quantum mechanics: one may formulate an eigenvalue problem for the electric (or magnetic) field and construct the corresponding right and left eigenstates using the standard quantum-mechanical inner product~\cite{truong2020dqnm}. Within such a formulation, the left eigenstates have the same character as the right eigenstates, consisting solely of electric (or magnetic) fields.

The biorthogonality relation associated with the full Maxwell equations, however, is known to take a fundamentally different form. In particular, unconjugated Lorentz reciprocity, which applies naturally to systems with gain and loss, yields a biorthogonality relation involving both the electric and magnetic fields and an integration structure that differs markedly from the standard quantum-mechanical inner product~\cite{bresler1958,snyder1983,collin1991,vassallo1991}. While this relation has long provided a practical foundation for modal expansions in electromagnetic systems, its connection to the conventional notion of left and right eigenstates is not immediately apparent. This apparent mismatch raises a fundamental question: how should the left and right eigenstates of the full Maxwell equations be properly defined and interpreted, and how does their biorthogonal structure relate to that obtained from Lorentz reciprocity? Figure~\ref{fig:Schematic} outlines these questions schematically.

In this work, we revisit the notions of left and right eigenstates for electromagnetic waves in non-Hermitian photonic systems directly from the full vectorial Maxwell equations, without invoking paraxial or coupled-mode approximations. More specifically, in this work, we focus on the guided modes of non-Hermitian waveguides—that is, spatially bound states, which may have complex propagation constants—and do not consider radiation modes or open cavities supporting quasinormal modes~\cite{sauvan2013qnm,Ge_2014,kristensen2020qnm,sauvan2022qnm}. Starting from the electric-field wave equation for waveguiding structures, we derive the corresponding right and left eigenvalue problems and establish their biorthogonality. We then connect this eigenstate formulation to unconjugated Lorentz reciprocity and show how the biorthogonality relation obtained from an electric-field-only (or magnetic-field-only) wave equation is reconciled with the reciprocity relation involving both electric and magnetic fields. Rather than introducing a new orthogonality relation, our analysis establishes the operator-theoretic connection between these two formulations, providing a unified interpretation of left and right eigenstates and electromagnetic biorthogonality within the full vectorial Maxwell framework.

\section{Biorthogonality in Scalar Optics}
We start by recalling how the concept of biorthogonality has traditionally been applied in optics. Much of the work in non-Hermitian optics has focused on coupled systems, such as waveguides or cavities, that can be described within the framework of spatial or temporal coupled-mode theory. For example, in spatial coupled-mode theory the system is described by
\begin{equation}
i\frac{d\vec{a}}{dz}= M \vec{a},
\label{Eq.CMT}
\end{equation}
where $\vec{a}$ is a field-amplitude vector whose element $a_n$ represents the field amplitude at site $n$, $M$ is the system matrix (often denoted by $H$, a notation we avoid here to prevent confusion with the magnetic field), and $z$ is the propagation distance. In temporal coupled-mode theory, $z$ is replaced by time $t$, and additional terms describing the input (excitation) and output channels are included. These additional terms, however, do not affect the eigenmodes of the system.

In non-Hermitian optics, $M$ is generally a non-normal matrix, i.e., $[M,M^{\dagger}] \neq 0$, where $[\,\cdot\,,\,\cdot\,]$ denotes the commutator and the superscript $\dagger$ denotes the Hermitian adjoint (transpose and complex conjugation). In this case, the right eigenvectors of $M$, defined by $M\vec{v}_n=\mu_n\vec{v}_n$, are not necessarily orthogonal. Nevertheless, constructing a biorthogonal basis is extremely useful in many applications, particularly for decomposing an arbitrary field vector in terms of the eigenstates, i.e., $\vec{V}=\sum_n c_n\vec{v}_n$. The standard procedure is therefore to introduce the left eigenvectors, defined by $\vec{u}_n^{\dagger}M=\mu_n\vec{u}_n^{\dagger}$. Projecting the equation for the right eigenvectors onto $\vec{u}_m^{\dagger}$ gives $\vec{u}_m^{\dagger}M\vec{v}_n=\mu_n\vec{u}_m^{\dagger}\vec{v}_n$. Similarly, projecting the equation for the left eigenvectors onto $\vec{v}_n$ yields $\vec{u}_m^{\dagger}M\vec{v}_n=\mu_m\vec{u}_m^{\dagger}\vec{v}_n$. Subtracting these two relations leads to $(\mu_n-\mu_m)\vec{u}_m^{\dagger}\vec{v}_n=0$. Therefore, for distinct eigenvalues, i.e., $\mu_n\neq\mu_m$, one obtains the biorthogonality relation $\vec{u}_m^{\dagger}\vec{v}_n=0$, with the normalization convention typically chosen such that $\vec{u}_m^{\dagger}\vec{v}_n=\delta_{mn}$.

Importantly, the equation for the left eigenvectors can be rewritten as
$M^{\dagger}\vec{u}_n=\mu_n^*\vec{u}_n$, where the superscript $*$ denotes complex conjugation. This form explicitly highlights the role of the adjoint operator. While this observation may appear to be of limited significance for finite-dimensional matrix problems, it becomes crucial when dealing with continuous non-Hermitian operators. To illustrate this point, consider a continuous non-Hermitian system described by a Schr\"odinger-like equation. Such equations played a central role in the early development of $\mathcal{PT}$-symmetric quantum mechanics and also arise naturally in optics through the scalar paraxial wave equation, which is mathematically analogous to the single-particle Schr\"odinger equation:
\begin{equation}
i\frac{d\ket{\psi}}{dz}= \mathcal{L}\ket{\psi},
\label{Eq.Schrodinger}
\end{equation}
where $\mathcal{L}=-\nabla_t^2+V(x,y)$, with $\nabla_t^2$ denoting the transverse Laplacian relevant to optical waveguide problems. Bound modes of this equation are defined as those whose fields decay exponentially outside the waveguide core and vanish at infinity. The right eigenfunctions of $\mathcal{L}$ are defined by $\mathcal{L}\ket{\phi_n}=\mu_n\ket{\phi_n}$.

In practice, the biorthogonal basis can be obtained by discretizing the operator and solving the resulting matrix eigenvalue problem, thereby reducing Eq.~(\ref{Eq.Schrodinger}) to the form of Eq.~(\ref{Eq.CMT}). From a formal perspective, however, it is desirable to construct the biorthogonal basis directly within the continuous description. This requires first defining the adjoint of a continuous operator. Formally, $\mathcal{L}^{\dagger}$ is the adjoint of $\mathcal{L}$ if $\braket{\psi_2|\mathcal{L}\psi_1}=\braket{\mathcal{L}^{\dagger}\psi_2|\psi_1}$ for all admissible states $\ket{\psi_1}$ and $\ket{\psi_2}$.

Let us now assume that the adjoint operator admits the eigenvalue problem $\mathcal{L}^{\dagger}\ket{\eta_n}=l_n\ket{\eta_n}$. Projecting the eigenvalue equation for $\mathcal{L}$ onto $\bra{\eta_m}$ yields $\braket{\eta_m|\mathcal{L}\phi_n}=\mu_n\braket{\eta_m|\phi_n}$. Using the definition of the adjoint operator, this relation can be rewritten as $\braket{\mathcal{L}^{\dagger}\eta_m|\phi_n}=\mu_n\braket{\eta_m|\phi_n}$. On the other hand, projecting the eigenvalue equation for $\mathcal{L}^{\dagger}$ onto $\bra{\phi_n}$ gives $\braket{\phi_n|\mathcal{L}^{\dagger}\eta_m}=l_m\braket{\phi_n|\eta_m}$. Taking the complex conjugate of this expression yields $\braket{\mathcal{L}^{\dagger}\eta_m|\phi_n}=l_m^*\braket{\eta_m|\phi_n}$. Combining these two relations then gives $(\mu_n-l_m^*)\braket{\eta_m|\phi_n}=0$.

If we consider only the subspace spanned by the discrete bound modes and assume that the sets $\{\phi_n\}$ and $\{\eta_m\}$ form complete bases within this subspace. Under this condition, for each eigenfunction $\ket{\eta_m}$ there must exist at least one eigenfunction $\ket{\phi_n}$ such that $\braket{\eta_m|\phi_n}\neq 0$; otherwise, $\ket{\eta_m}$ would be orthogonal to the entire basis $\{\phi_n\}$ and hence must vanish. We assign the same index to eigenfunctions with nonzero overlap, such that $\braket{\eta_n|\phi_n}\neq 0$. It then follows from the relation derived above that $\mu_n=l_n^*$. Assuming a nondegenerate spectrum, $\mu_n\neq l_m^*$ for $n\neq m$, and therefore $\braket{\eta_m|\phi_n}=0$ whenever $n\neq m$. This establishes the correspondence between the eigenfunctions of $\mathcal{L}$ and those of its adjoint $\mathcal{L}^{\dagger}$ and shows that the two sets form a biorthogonal basis.

\section{Biorthogonality in the Full-Vectorial Maxwell Eigenproblem}\label{sec:biorthvect}
\noindent
\textbf{Lorentz Reciprocity and Standard Modal Orthogonality:---} Having reviewed the concept of biorthogonality within the scalar paraxial and coupled-mode frameworks commonly used in non-Hermitian optics, we now turn to the full vectorial Maxwell equations. Specifically, we consider guided modes of non-Hermitian waveguiding structures containing spatially distributed gain and loss. Our goal is to establish the corresponding left and right electromagnetic eigenstates and derive their biorthogonality relations directly from the full Maxwell eigenvalue problem. Such a formulation is essential when scalar or weak-guidance approximations are not applicable and provides a rigorous foundation for modal expansions and coupled-mode descriptions of vectorial non-Hermitian electromagnetic systems.

From the previous discussion, one might expect that a biorthogonality relation similar to that introduced above could be constructed for the full vectorial electromagnetic problem. After all, for a step-index waveguide, one typically writes the wave equation separately in each homogeneous layer (core and cladding), solves for the fields in each region, and then applies the appropriate boundary conditions to obtain the guided modes. In each layer, the electric-field wave equation takes the form $\nabla_t^2 \mathbf{E}_j + k_0^2 \epsilon_j \mathbf{E}_j = \beta^2 \mathbf{E}_j$, where $\mathbf{E}_j$ is the electric-field vector in layer $j$, $\epsilon_j$ is the corresponding relative permittivity, $\beta$ is the propagation constant, $k_0 = \omega/c$ is the free space wavenumber, $\omega$ is the angular frequency, and $c$ is the speed of light.

Comparing this equation with the eigenvalue equation introduced above for paraxial optics through the operator $\mathcal{L}$, one might conclude that the two problems have essentially the same mathematical structure, apart from the fact that the electromagnetic wave equation describes a vector field rather than a scalar field. The appearance of the eigenvalue as $\beta^2$ rather than $\beta$ poses no fundamental difficulty, since one may simply define $\lambda=\beta^2$. Of course, the boundary conditions differ depending on the waveguiding structure and, in particular, on whether the waveguide is dielectric or metallic, but this does not appear to alter the basic eigenvalue structure.

Following this reasoning, one might then expect that the left eigenfunctions of the full vectorial problem could be obtained simply by constructing the adjoint of the corresponding electric-field operator, thereby yielding a biorthogonal basis in direct analogy with the scalar paraxial case. Both the right eigenfunctions and their biorthogonal partners would then be expressed entirely in terms of the electric field and related through the same type of inner product employed above. An analogous construction could, in principle, be carried out starting from the magnetic-field wave equation.

However, the standard orthogonality relation for electromagnetic guided modes, derived from unconjugated Lorentz reciprocity, takes a markedly different form \cite{bresler1958,snyder1983,collin1991}. Consider two guided modes, labeled $m$ and $n$, with transverse field profiles $\{\mathbf{e}_m,\mathbf{h}_m\}$ and $\{\mathbf{e}_n,\mathbf{h}_n\}$ and propagation constants $\beta_m$ and $\beta_n$, respectively. Their full fields vary along the waveguide as $\mathbf{E}_m(x,y,z)=\mathbf{e}_m(x,y)e^{i\beta_m z}$, $\mathbf{H}_m(x,y,z)=\mathbf{h}_m(x,y)e^{i\beta_m z}$, and similarly for mode $n$. Applying unconjugated Lorentz reciprocity gives
\[
(\beta_m+\beta_n)
\int_\Omega
\left(
\mathbf{e}_m\times\mathbf{h}_n
-
\mathbf{e}_n\times\mathbf{h}_m
\right)\cdot\hat{\mathbf{z}}\,d\Omega=0,
\]
where $\Omega$ denotes the transverse cross section of the waveguide and $\hat{\mathbf{z}}$ is the unit vector along the propagation direction. Consequently, if $\beta_m+\beta_n\neq0$, the two modes are orthogonal under the bilinear form appearing above. A nonzero overlap can occur only for mode pairs satisfying $\beta_m=-\beta_n$, corresponding to counterpropagating partners when the propagation constants are defined with sign. Unlike the biorthogonality relation suggested by the electric-field wave equation, this relation involves both the electric and magnetic fields and does not employ the standard Hermitian inner product. We note that the  standard derivation of the above relation is algebraic and does not invoke operator theory or the concepts of left and right eigenstates. For completeness, we provide this derivation, together with that of the conjugated Lorentz reciprocity relation, in Appendix~\ref{app:lorentz}. 

\begin{figure}[t]
    \centering
    \includegraphics[width=\columnwidth]{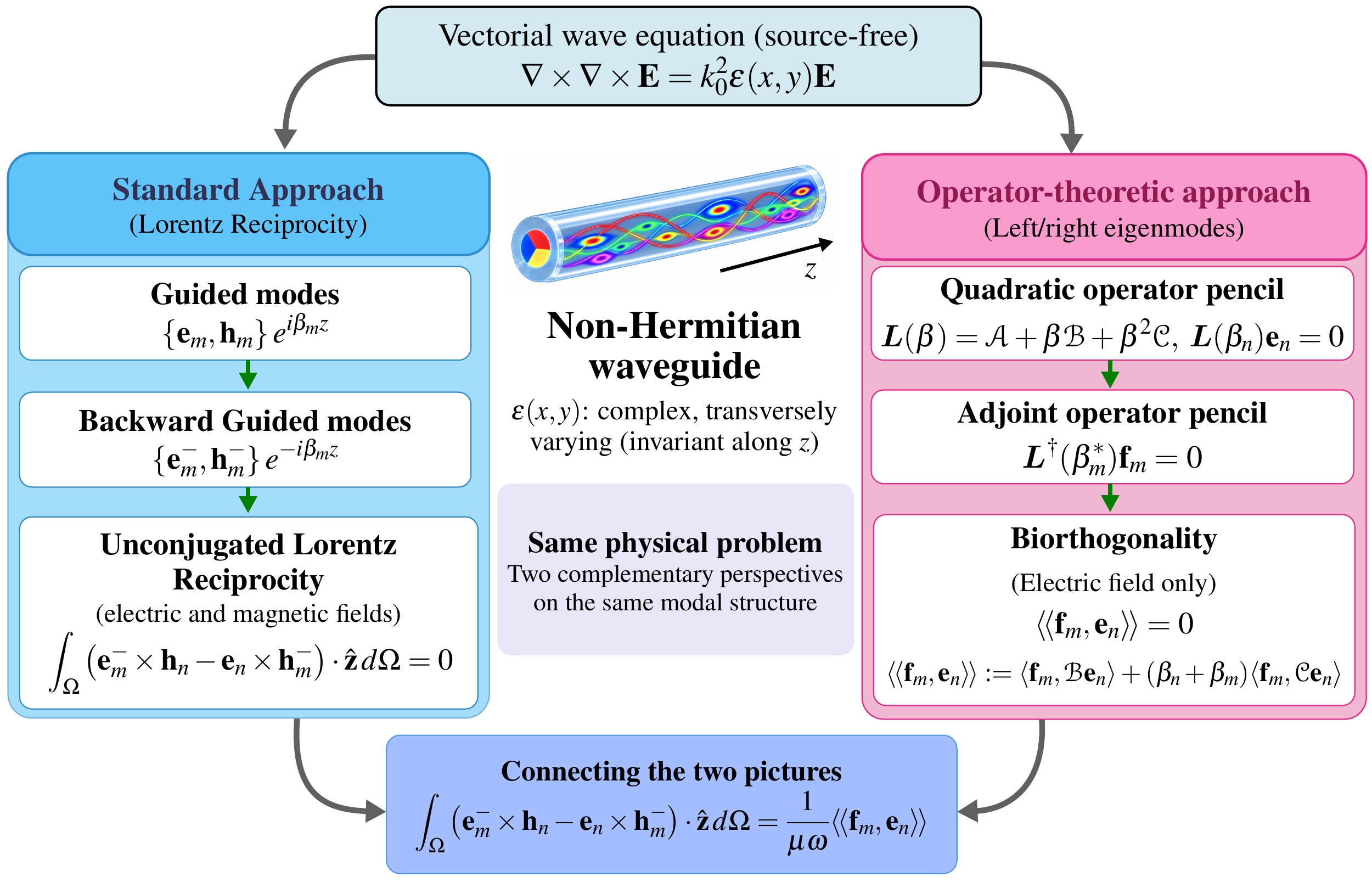}
    \caption{Schematic overview of the two approaches to modal orthogonality in non-Hermitian electromagnetic waveguides. The standard approach, based on unconjugated Lorentz reciprocity, leads to an orthogonality relation involving both the electric and magnetic fields. The operator-theoretic approach formulates the vectorial Maxwell wave equation as a quadratic operator pencil and introduces the corresponding left and right eigenmodes and their biorthogonality. The two formulations are connected through the relation shown at the bottom, establishing their equivalence.}
    \label{fig:Workflow}
\end{figure}

This apparent difference between the two formulations naturally raises the question of how they are related: on the one hand, the well-established electromagnetic biorthogonality relations derived from Lorentz reciprocity and, on the other hand, the framework based on adjoint operators and right and left eigenfunctions. In what follows, we establish the precise connection between these two perspectives. To guide the discussion and clarify the connection between these two perspectives, Fig. \ref{fig:Workflow} summarizes the two routes to electromagnetic biorthogonality---one based on adjoint operators and left/right eigenstates, and the other on Lorentz reciprocity---and the correspondence between them established below.\\

\noindent
\textbf{Operator-Theoretic Derivation of Electromagnetic Left Eigenstates:---} 
We begin by noting that the electric-field wave equation written in terms of the operator $\nabla_t^2+k_0^2\epsilon_j$ applies within a homogeneous region characterized by a constant permittivity $\epsilon_j$ and therefore does not, by itself, provide a global eigenvalue equation for the electric field of a guided mode across the entire waveguide cross section. The appearance of the Laplacian follows from the vector identity $\nabla\times\nabla\times\mathbf{E}=\nabla(\nabla\cdot\mathbf{E})-\nabla^2\mathbf{E}$ together with the condition $\nabla\cdot\mathbf{E}=0$, which holds within a homogeneous source-free dielectric region. When the complete waveguide structure is considered, however, the permittivity varies across the transverse plane, and $\nabla\cdot\mathbf{E}=0$ does not generally hold throughout the entire domain. Instead, in the absence of free charge, Maxwell's equations require $\nabla\cdot[\epsilon(\mathbf{r})\mathbf{E}]=0$.

A global description of the guided modes must therefore start from the full vectorial electric-field wave equation,
\begin{equation}
\nabla\times\nabla\times\mathbf{E}-k_0^2\epsilon(x,y)\mathbf{E}=0,
\label{eq:waveeq}
\end{equation}
where $\epsilon(x,y)$ is the relative permittivity, which may be complex to describe gain or loss. Throughout, we adopt the $e^{-i\omega t}$ time convention, such that $\mathrm{Im}\,\epsilon>0$ corresponds to loss and $\mathrm{Im}\,\epsilon<0$ to gain, and assume nonmagnetic, isotropic, and $z$-invariant media with permeability $\mu=\mu_0$. Throughout this work, we focus on guided (bound) modes whose fields decay at infinity, ensuring that all boundary terms arising from the integrations by parts performed below vanish.

We proceed by seeking guided-mode solutions with a well-defined propagation constant $\beta$ along the $z$ direction. Since the waveguide is invariant along $z$, the electric field can be written as
\begin{equation}
\mathbf{E}(x,y,z)=\mathbf{e}(x,y)e^{i\beta z},
\label{eq:modal_ansatz}
\end{equation}
where $\mathbf{e}(x,y)$ is the vectorial modal profile in the transverse plane. The three-dimensional gradient operator can be decomposed as
\begin{equation}
\nabla=\nabla_t+\hat{\mathbf{z}}\frac{\partial}{\partial z},
\end{equation}
where
\begin{equation}
\nabla_t=\hat{\mathbf{x}}\frac{\partial}{\partial x}
+\hat{\mathbf{y}}\frac{\partial}{\partial y}
\end{equation}
contains only derivatives with respect to the transverse coordinates. Because $\mathbf{e}(x,y)$ is independent of $z$, differentiation of the modal field with respect to $z$ gives
\begin{equation}
\frac{\partial}{\partial z}
\left[\mathbf{e}(x,y)e^{i\beta z}\right]
=
i\beta\,\mathbf{e}(x,y)e^{i\beta z}.
\end{equation}
Therefore, whenever $\nabla$ acts on a field of the form given in Eq.~\eqref{eq:modal_ansatz}, its action can be written as $\nabla \longrightarrow \nabla_t+i\beta\hat{\mathbf{z}}$
after the common factor $e^{i\beta z}$ has been removed.

Substituting the modal ansatz into Eq.~\eqref{eq:waveeq} therefore gives
\begin{equation}
\left[
\left(\nabla_t+i\beta\hat{\mathbf{z}}\right)
\times
\left(\nabla_t+i\beta\hat{\mathbf{z}}\right)
\times
-k_0^2\epsilon(x,y)
\right]\mathbf{e}=0.
\label{eq:waveeq_beta}
\end{equation}
Expanding the two curl operators explicitly, we obtain
\begin{align}
&\nabla_t\times\left(\nabla_t\times\mathbf{e}\right)
+i\beta\,\nabla_t\times\left(\hat{\mathbf{z}}\times\mathbf{e}\right)
+i\beta\,\hat{\mathbf{z}}\times\left(\nabla_t\times\mathbf{e}\right)
\nonumber\\
&\qquad
-\beta^2\hat{\mathbf{z}}\times
\left(\hat{\mathbf{z}}\times\mathbf{e}\right)
-k_0^2\epsilon(x,y)\mathbf{e}=0.
\label{eq:expanded_beta}
\end{align}
The different terms in this equation can now be grouped according to their dependence on the propagation constant $\beta$. Defining
\begin{equation}
\A\mathbf{e}
\equiv
\nabla_t\times\left(\nabla_t\times\mathbf{e}\right)
-k_0^2\epsilon(x,y)\mathbf{e},
\label{eq:Aoperator}
\end{equation}
\begin{equation}
\B\mathbf{e}
\equiv
i\left[
\nabla_t\times\left(\hat{\mathbf{z}}\times\mathbf{e}\right)
+
\hat{\mathbf{z}}\times\left(\nabla_t\times\mathbf{e}\right)
\right],
\label{eq:Boperator}
\end{equation}
and
\begin{equation}
\C\mathbf{e}
\equiv
-\hat{\mathbf{z}}\times
\left(\hat{\mathbf{z}}\times\mathbf{e}\right),
\label{eq:Coperator}
\end{equation}
Eq.~\eqref{eq:expanded_beta} becomes
\begin{equation}
\left(\A+\beta \B+\beta^2 \C\right)\mathbf{e}=0.
\label{eq:quadpencil}
\end{equation}
Thus, unlike the scalar paraxial problem discussed previously, the full vectorial Maxwell equation does not lead directly to a conventional linear eigenvalue problem in $\beta$. Instead, it gives a quadratic operator eigenvalue problem, or equivalently a quadratic operator pencil,
\begin{equation}
\boldsymbol{L}(\beta)
\equiv
\A+\beta \B+\beta^2\C,
\end{equation}
with the guided modes determined by
\begin{equation}
\boldsymbol{L}(\beta)\mathbf{e}=0.
\end{equation}

To make the structure of the operators $\A$, $\B$, and $\C$ more explicit, we decompose the modal electric field into its transverse and longitudinal components,
\begin{equation}
\mathbf{e}=\mathbf{e}_t+\hat{\mathbf{z}}e_z,
\end{equation}
where $\mathbf{e}_t=e_x\hat{\mathbf{x}}+e_y\hat{\mathbf{y}}$ and $e_z$ is the longitudinal field component. By using standard vector identities, we can show that (see Appendix~\ref{app:pencilEquations}):
\begin{equation}
\A\mathbf{e}
=
\mathcal{T}\mathbf{e}_t
-\hat{\mathbf{z}}\mathcal{S}e_z,
\end{equation}
where
\begin{equation}
\mathcal{T}\equiv\nabla_t\times\nabla_t\times-k_0^2\epsilon,
\qquad
\mathcal{S}\equiv\nabla_t^2+k_0^2\epsilon.
\end{equation}

Similarly for the operators $\B$ and $\C$, we obtain:
\begin{equation}
\B\mathbf{e}
=
i\left[
\nabla_t e_z
+
\hat{\mathbf{z}}(\nabla_t\cdot\mathbf{e}_t)
\right],
\end{equation}
and
\begin{equation}
\C\mathbf{e}=\mathbf{e}_t.
\end{equation}

To facilitate the subsequent analysis and make the underlying symmetry properties of the eigenvalue problem explicit, we separate the quadratic pencil equation into its transverse and longitudinal components. The resulting coupled equations contain both linear and quadratic dependence on the propagation constant $\beta$. To recast these equations into a generalized eigenvalue problem (GEVP) with eigenvalue $\lambda=\beta^2$, we introduce the rescaled longitudinal field $\tilde{e}_z=i\beta e_z$. With this transformation, the coupled equations can be written in the $2\times2$ block form:
\begin{equation}
\begin{bmatrix}
-\mathcal{T} & -\nabla_t \\[3pt]
0 & \mathcal{S}
\end{bmatrix}
\begin{bmatrix}
\mathbf{e}_t\\
\tilde{e}_z
\end{bmatrix}
=
\lambda
\begin{bmatrix}
I & 0\\[3pt]
-\nabla_t\cdot & 0
\end{bmatrix}
\begin{bmatrix}
\mathbf{e}_t\\
\tilde{e}_z
\end{bmatrix},
\label{eq:block_gevp}
\end{equation}
as derived explicitly in Appendix~\ref{app:gevp}.
An important consequence of the transformation $\tilde e_z=i\beta e_z$ is that the propagation constant appears in Eq.~\eqref{eq:block_gevp} only through $\lambda=\beta^2$. Therefore, if a particular pair $(\mathbf{e}_t,\tilde e_z)$ is an eigenvector of the GEVP with eigenvalue $\lambda=\beta^2$, exactly the same eigenvector is associated with the two possible propagation constants $+\beta$ and $-\beta$. In other words, changing from one root of $\beta^2$ to the other does not change either $\mathbf{e}_t$ or $\tilde e_z$ in the block eigenvalue problem.

The physical longitudinal electric-field component, however, is not $\tilde e_z$, but is recovered from the transformation used to construct the GEVP as
\begin{equation}
e_z=\frac{\tilde e_z}{i\beta}.
\end{equation}
For the $+\beta$ root, we therefore have
\begin{equation}
e_z^{+}=\frac{\tilde e_z}{i\beta},
\end{equation}
whereas for the $-\beta$ root, the same $\tilde e_z$ gives
\begin{equation}
e_z^{-}
=\frac{\tilde e_z}{i(-\beta)}
=-e_z^{+}.
\end{equation}
Thus, the sign reversal of $e_z$ does not occur within the block eigenvalue problem itself. Rather, it appears only when the common GEVP eigenvector $(\mathbf{e}_t,\tilde e_z)$ is converted back to the physical electric-field components.

Consequently, the two roots $\pm\beta$ correspond to the physical transverse profiles
\begin{equation}\label{eq:pm_modes}
\mathbf{e}^{+}
=
\mathbf{e}_t+\hat{\mathbf{z}}e_z,
\qquad
\mathbf{e}^{-}
=
\mathbf{e}_t-\hat{\mathbf{z}}e_z,
\end{equation}
respectively. 
The corresponding full electric fields for the two propagation directions can therefore be written as
\begin{align}
\mathbf{E}^{+}(x,y,z)
&=
\left(\mathbf{e}_t+\hat{\mathbf{z}}e_z\right)e^{i\beta z},\\
\mathbf{E}^{-}(x,y,z)
&=
\left(\mathbf{e}_t-\hat{\mathbf{z}}e_z\right)e^{-i\beta z}.
\end{align}
Thus, the two counter-propagating solutions have identical transverse electric-field components and opposite longitudinal components. This is the familiar relation between forward- and backward-propagating guided modes, here recovered directly from the block generalized eigenvalue formulation.

Having formulated the Maxwell problem as the quadratic eigenvalue equation
\begin{equation}
\boldsymbol{L}(\beta)\mathbf{e}
=
\left(\A+\beta \B+\beta^2\C\right)\mathbf{e}
=0,
\end{equation}
we now seek to construct the corresponding left eigenmodes and determine their orthogonality relation with the right eigenmodes. The procedure parallels the finite-dimensional discussion of Sec. 2: we first construct the adjoint of the operator defining the right eigenvalue problem, use it to define the left eigenvalue problem, and then combine the right and left eigenmode equations to obtain the corresponding biorthogonality relation. The main difference here is that the Maxwell eigenvalue problem depends quadratically on the propagation constant $\beta$, and this modifies the form of the resulting orthogonality relation.

To construct the adjoint problem, we adopt the standard sesquilinear inner product over the transverse waveguide cross section $\Omega$,
\begin{equation}
\langle\mathbf{f},\mathbf{e}\rangle
=
\int_\Omega\mathbf{f}^*\cdot\mathbf{e}\,d\Omega,
\end{equation}
and define the formal adjoint $X^\dagger$ of an operator $X$ through
\begin{equation}
\langle\mathbf{f},X\mathbf{e}\rangle
=
\langle X^\dagger\mathbf{f},\mathbf{e}\rangle.
\end{equation}
Here, the adjoint is understood in the formal sense: differential operators are transferred from one field to the other through integration by parts, with the boundary terms vanishing for the guided modes considered here.

We next apply this definition to the quadratic pencil $\boldsymbol{L}(\beta)$. Because the inner product is conjugate-linear in its first argument, the scalar coefficients multiplying the operators must be complex conjugated when transferred to the first argument. For example,
\begin{equation}
\langle\mathbf{f},\beta \B\mathbf{e}\rangle
=
\beta\langle\mathbf{f},\B\mathbf{e}\rangle
=
\langle\beta^*\B^\dagger\mathbf{f},\mathbf{e}\rangle,
\end{equation}
with an analogous relation for the term proportional to $\beta^2$. Applying this procedure to all three terms gives
\begin{equation}
\langle\mathbf{f},\boldsymbol{L}(\beta)\mathbf{e}\rangle
=
\left\langle
\left[\A^\dagger+\beta^*\B^\dagger+(\beta^*)^2\C^\dagger\right]
\mathbf{f},
\mathbf{e}
\right\rangle.
\end{equation}
It is therefore natural to define the adjoint pencil
\begin{equation}
\boldsymbol{L}^\dagger(\zeta)
=
\A^\dagger+\zeta \B^\dagger+\zeta^2\C^\dagger,
\end{equation}
so that
\begin{equation}
[\boldsymbol{L}(\beta)]^\dagger
=
\boldsymbol{L}^\dagger(\beta^*).
\label{eq:adjoint_pencil_identity}
\end{equation}

With the adjoint pencil established, we can now define the left eigenmode associated with a right eigenvalue $\beta_m$. In direct analogy with the usual non-Hermitian eigenvalue problem, the right and left eigenmodes satisfy
\begin{align}
\boldsymbol{L}(\beta_n)\mathbf{e}_n&=0,
\label{eq:right_pencil_mode}\\
\boldsymbol{L}^\dagger(\beta_m^*)\mathbf{f}_m&=0,
\label{eq:left_pencil_mode}
\end{align}
respectively. The explicit form of $\mathbf{f}_m$ will be determined below. At this stage, these two equations provide the right and left eigenvalue problems whose mutual orthogonality we wish to establish.

To obtain this orthogonality relation, we now compare the two eigenmode equations through the inner product. Taking the inner product of Eq.~\eqref{eq:right_pencil_mode} with $\mathbf{f}_m$ gives
\begin{equation}
\langle\mathbf{f}_m,\boldsymbol{L}(\beta_n)\mathbf{e}_n\rangle=0.
\label{eq:right_projected}
\end{equation}
On the other hand, using Eq.~\eqref{eq:adjoint_pencil_identity} together with the left eigenmode equation gives
\begin{equation}
\langle\mathbf{f}_m,\boldsymbol{L}(\beta_m)\mathbf{e}_n\rangle
=
\langle\boldsymbol{L}^\dagger(\beta_m^*)\mathbf{f}_m,\mathbf{e}_n\rangle
=0.
\label{eq:left_projected}
\end{equation}
We therefore have two relations involving the same pair of fields, $\mathbf{f}_m$ and $\mathbf{e}_n$, but with the pencil evaluated at the two different propagation constants $\beta_m$ and $\beta_n$. Subtracting them isolates precisely this difference:
\begin{equation}
\left\langle
\mathbf{f}_m,
\left[\boldsymbol{L}(\beta_n)-\boldsymbol{L}(\beta_m)\right]
\mathbf{e}_n
\right\rangle
=0.
\end{equation}

The quadratic dependence of $\boldsymbol{L}$ on $\beta$ now determines the form of the resulting orthogonality relation. In particular,
\begin{align}
\boldsymbol{L}(\beta_n)-\boldsymbol{L}(\beta_m)
&=
(\beta_n-\beta_m)\B
+
(\beta_n^2-\beta_m^2)\C\\
&=
(\beta_n-\beta_m)
\left[\B+(\beta_n+\beta_m)\C\right].
\end{align}
Consequently,
\begin{equation}
(\beta_n-\beta_m)
\left[
\langle\mathbf{f}_m,\B\mathbf{e}_n\rangle
+
(\beta_n+\beta_m)
\langle\mathbf{f}_m,\C\mathbf{e}_n\rangle
\right]
=0.
\label{eq:pencil_factorized}
\end{equation}
This result suggests defining the following pairing between a left eigenmode $\mathbf{f}_m$ and a right eigenmode $\mathbf{e}_n$:
\begin{equation}
\bilin{\mathbf{f}_m,\mathbf{e}_n}
:=
\langle\mathbf{f}_m,\B\mathbf{e}_n\rangle
+
(\beta_n+\beta_m)
\langle\mathbf{f}_m,\C\mathbf{e}_n\rangle.
\label{eq:bilinearF}
\end{equation}
Equation~\eqref{eq:pencil_factorized} can then be written simply as
\begin{equation}
(\beta_n-\beta_m)
\bilin{\mathbf{f}_m,\mathbf{e}_n}
=0.
\end{equation}
Therefore, for distinct propagation constants $\beta_n\neq\beta_m$,
\begin{equation}
\bilin{\mathbf{f}_m,\mathbf{e}_n}=0.
\label{eq:pencil_orthogonality}
\end{equation}

This result is the analogue, for the quadratic Maxwell eigenvalue problem, of the familiar biorthogonality relation obtained for a linear non-Hermitian eigenvalue problem. Importantly, the orthogonality here is not defined with respect to the original inner product $\langle\mathbf{f}_m,\mathbf{e}_n\rangle$. Rather, the quadratic dependence on $\beta$ leads naturally to the modified pairing in Eq.~\eqref{eq:bilinearF}, which contains contributions from both $\B$ and $\C$ and depends explicitly on the propagation constants $\beta_m$ and $\beta_n$.

We have thus established the biorthogonality relation associated with the quadratic Maxwell eigenvalue problem. The next step is to determine the physical electromagnetic field represented by the left eigenfunction $\mathbf{f}_m$ and to show how the pairing in Eq.~\eqref{eq:bilinearF} is related to the familiar electromagnetic orthogonality relation obtained from Lorentz reciprocity.

We now determine the physical field corresponding to the left eigenmode $\mathbf{f}_m$. This is an important step because the adjoint eigenvalue equation
\begin{equation}
\boldsymbol{L}^\dagger(\beta_m^*)\mathbf{f}_m=0,
\end{equation}
defines $\mathbf{f}_m$ mathematically, but does not yet reveal how this field is related to the physical electromagnetic modes of the original waveguide. We show below that $\mathbf{f}_m$ can be constructed directly from the backward-propagating partner of the corresponding right eigenmode.

As established above, a right eigenmode with propagation constant $\beta_m$ has a backward-propagating partner $\mathbf{e}_m^-$ with propagation constant $-\beta_m$. The latter therefore satisfies
\begin{equation}
\boldsymbol{L}(-\beta_m)\mathbf{e}_m^-=0,
\end{equation}
or, explicitly,
\begin{equation}
\left(\A-\beta_m \B+\beta_m^2 \C\right)\mathbf{e}_m^-=0.
\label{eq:backward_pencil}
\end{equation}
Our objective is to relate this equation to the adjoint eigenvalue equation at $\beta_m^*$.

For this purpose, the adjoint properties of the three operators entering the quadratic pencil are essential. As shown in Appendix~\ref{app:opid}, they satisfy
\begin{equation}
\A^*=\A^\dagger,
\qquad
\B^*=-\B^\dagger,
\qquad
\C^*=\C^\dagger.
\label{eq:operator_identities}
\end{equation}
We now complex conjugate Eq.~\eqref{eq:backward_pencil}, which gives
\begin{equation}
\left(
\A^*
-\beta_m^*\B^*
+(\beta_m^*)^2\C^*
\right)
(\mathbf{e}_m^-)^*
=0.
\end{equation}
Substituting the operator identities in Eq.~\eqref{eq:operator_identities} yields
\begin{align}
\left[
\A^\dagger
-\beta_m^*(-\B^\dagger)
+(\beta_m^*)^2\C^\dagger
\right]
(\mathbf{e}_m^-)^*
&=0\\
\Longrightarrow\qquad
\left[
\A^\dagger
+\beta_m^*\B^\dagger
+(\beta_m^*)^2\C^\dagger
\right]
(\mathbf{e}_m^-)^*
&=0.
\end{align}
The operator in square brackets is precisely the adjoint pencil $\boldsymbol{L}^\dagger(\beta_m^*)$. Hence,
\begin{equation}
\boldsymbol{L}^\dagger(\beta_m^*)
(\mathbf{e}_m^-)^*
=0.
\label{eq:adjoint_construct}
\end{equation}

Comparison with the defining equation for the left eigenmode shows that the complex conjugate of the backward-propagating electric field provides the required left eigenmode. We may therefore choose
\begin{equation}
\mathbf{f}_m
=
(\mathbf{e}_m^-)^*.
\label{eq:adjoint_identity}
\end{equation}
Using the relation between the forward- and backward-propagating electric-field profiles derived above:
\begin{equation}
\mathbf{e}_m^-
=
\mathbf{e}_{m,t}
-
\hat{\mathbf{z}}e_{m,z},
\end{equation}
the left eigenmode can be written entirely in terms of the forward-mode field components as
\begin{equation}
\boxed{
\mathbf{f}_m
=
\mathbf{e}_{m,t}^*
-
\hat{\mathbf{z}}e_{m,z}^*
}.
\label{eq:left_explicit}
\end{equation}

Thus, the left eigenmode of the full vectorial Maxwell problem is not an independent or abstract field that must be determined separately. It is obtained directly from the corresponding right eigenmode by first forming its backward-propagating partner, which reverses the sign of the longitudinal electric-field component, and then taking its complex conjugate. This explicit identification provides the connection needed to express the pencil biorthogonality relation in terms of the physical electric and magnetic fields and, as shown next, to relate it to the familiar orthogonality relation obtained from Lorentz reciprocity.\\

\noindent
\textbf{From Maxwell Biorthogonality to Lorentz Reciprocity:---} 
Having established the left eigenmode associated with the quadratic Maxwell eigenvalue problem, we now connect the resulting pencil biorthogonality relation to the conventional modal orthogonality relation derived from unconjugated Lorentz reciprocity. The two relations appear different at first sight: the pencil pairing is expressed entirely in terms of the electric-field eigenfunctions, whereas the Lorentz-reciprocity relation involves both electric and magnetic fields. To compare them directly, we first express the magnetic-field components in terms of the electric field using Maxwell's equations.

For the $e^{-i\omega t}$ convention adopted here, Faraday's law is
\begin{equation}
\nabla\times\mathbf{E}
=
i\omega\mu\mathbf{H}.
\label{eq:faraday_bridge}
\end{equation}
For a guided mode with propagation constant $\beta$,
\begin{equation}
\mathbf{E}(x,y,z)
=
\left(
\mathbf{e}_t+\hat{\mathbf{z}}e_z
\right)e^{i\beta z},
\end{equation}
we use $\nabla=\nabla_t+\hat{\mathbf{z}}\partial_z$ to obtain
\begin{align}
\nabla\times\mathbf{E}
&=
\left(\nabla_t+\hat{\mathbf{z}}\partial_z\right)
\times
\left[
\left(\mathbf{e}_t+\hat{\mathbf{z}}e_z\right)e^{i\beta z}
\right]\\
&=
e^{i\beta z}
\left[
\nabla_t\times\mathbf{e}_t
+
\nabla_t\times(\hat{\mathbf{z}}e_z)
+
i\beta\hat{\mathbf{z}}\times\mathbf{e}_t
\right].
\end{align}
The first term is directed along $\hat{\mathbf{z}}$, whereas the remaining two terms are transverse. Therefore, the transverse part of Faraday's law gives
\begin{equation}
i\omega\mu\,\mathbf{h}_t
=
\nabla_t\times(\hat{\mathbf{z}}e_z)
+
i\beta\hat{\mathbf{z}}\times\mathbf{e}_t.
\end{equation}
Since
\begin{equation}
\nabla_t\times(\hat{\mathbf{z}}e_z)
=
\nabla_t e_z\times\hat{\mathbf{z}}
=
-\hat{\mathbf{z}}\times\nabla_t e_z,
\end{equation}
we obtain
\begin{equation}
\mathbf{h}_t
=
\frac{1}{i\omega\mu}
\hat{\mathbf{z}}\times
\left(
i\beta\mathbf{e}_t-\nabla_t e_z
\right).
\label{eq:ht_from_e}
\end{equation}
This relation allows the magnetic field appearing in the Lorentz-reciprocity expression to be written entirely in terms of the electric-field eigenfunction.

We now apply Eq.~\eqref{eq:ht_from_e} to the two modes that enter the reciprocity relation. For the mode $\mathbf{e}_n$ with propagation constant $\beta_n$,
\begin{equation}
\mathbf{h}_{n,t}
=
\frac{1}{i\omega\mu}
\hat{\mathbf{z}}\times
\left(
i\beta_n\mathbf{e}_{n,t}
-
\nabla_t e_{n,z}
\right).
\label{eq:ht_n}
\end{equation}
For the backward-propagating partner of mode $m$, we have
\begin{equation}
\mathbf{e}_m^-
=
\mathbf{e}_{m,t}
-
\hat{\mathbf{z}}e_{m,z},
\qquad
\beta_m^-= -\beta_m.
\end{equation}
Substituting these quantities into Eq.~\eqref{eq:ht_from_e} gives
\begin{align}
\mathbf{h}_{m,t}^-
&=
\frac{1}{i\omega\mu}
\hat{\mathbf{z}}\times
\left(
-i\beta_m\mathbf{e}_{m,t}
+
\nabla_t e_{m,z}
\right)\\
&=
-\frac{1}{i\omega\mu}
\hat{\mathbf{z}}\times
\left(
i\beta_m\mathbf{e}_{m,t}
-
\nabla_t e_{m,z}
\right)
=
-\mathbf{h}_{m,t}.
\label{eq:ht_backward}
\end{align}
Thus, as expected for the counter-propagating pair, the transverse magnetic-field component changes sign.

We can now turn to the Lorentz-reciprocity expression. For the backward-propagating partner of mode $m$ and the forward-propagating mode $n$, define
\begin{equation}
\mathcal{F}_{mn}
=
\int_\Omega
\left(
\mathbf{e}_m^-\times\mathbf{h}_n
-
\mathbf{e}_n\times\mathbf{h}_m^-
\right)
\cdot\hat{\mathbf{z}}\,d\Omega.
\label{eq:lorentz_flux}
\end{equation}
Only the transverse components of the fields contribute to the longitudinal component of each cross product. Using
$\mathbf{e}_{m,t}^-=\mathbf{e}_{m,t}$ and
$\mathbf{h}_{m,t}^-=-\mathbf{h}_{m,t}$, Eq.~\eqref{eq:lorentz_flux} becomes
\begin{equation}
\mathcal{F}_{mn}
=
\int_\Omega
\left(
\mathbf{e}_{m,t}\times\mathbf{h}_{n,t}
+
\mathbf{e}_{n,t}\times\mathbf{h}_{m,t}
\right)
\cdot\hat{\mathbf{z}}\,d\Omega.
\label{eq:lorentz_flux_transverse}
\end{equation}

At this point, both magnetic fields in Eq.~\eqref{eq:lorentz_flux_transverse} can be eliminated using Eq.~\eqref{eq:ht_from_e}. Substituting their electric-field expressions, using the identification
\begin{equation}
\mathbf{f}_m
=
(\mathbf{e}_m^-)^*,
\end{equation}
derived in the preceding subsection, and performing one integration by parts gives
\begin{equation}
\mathcal{F}_{mn}
=
\frac{1}{\omega\mu}
\left[
\langle\mathbf{f}_m,\B\mathbf{e}_n\rangle
+
(\beta_n+\beta_m)
\langle\mathbf{f}_m,\C\mathbf{e}_n\rangle
\right],
\label{eq:bridge_expanded}
\end{equation}
where the intermediate algebra is given in Appendix~\ref{app:bridge}. The expression in square brackets is exactly the pencil pairing defined in Eq.~\eqref{eq:bilinearF}. We therefore arrive at
\begin{equation}
\boxed{
\mathcal{F}_{mn}
=
\frac{1}{\omega\mu}
\bilin{\mathbf{f}_m,\mathbf{e}_n}
}.
\label{eq:bridge}
\end{equation}

Equation~\eqref{eq:bridge} provides the direct connection between the two formulations. The left-hand side is the mixed electric--magnetic field expression appearing in the conventional Lorentz-reciprocity treatment, whereas the right-hand side is the electric-field-only pairing obtained from the adjoint structure of the quadratic Maxwell eigenvalue problem. The two expressions are therefore related exactly through Maxwell's equations.

The equivalence of the corresponding orthogonality relations follows immediately. For distinct propagation constants, $\beta_n\neq\beta_m$, the pencil biorthogonality relation derived above gives
\begin{equation}
\bilin{\mathbf{f}_m,\mathbf{e}_n}=0.
\end{equation}
Equation~\eqref{eq:bridge} then gives
\begin{equation}
\int_\Omega
\left(
\mathbf{e}_m^-\times\mathbf{h}_n
-
\mathbf{e}_n\times\mathbf{h}_m^-
\right)
\cdot\hat{\mathbf{z}}\,d\Omega
=
0,
\qquad
\beta_n\neq\beta_m.
\label{eq:lorentzorth}
\end{equation}
Thus, the modal orthogonality relation derived from unconjugated Lorentz reciprocity and the biorthogonality relation derived from the adjoint Maxwell eigenvalue problem are equivalent representations of the same modal orthogonality.

Having established the orthogonality of modes with distinct propagation constants, we next consider the diagonal case $n=m$. This case determines the self-pairing, or mode normalization,
\begin{equation}
s_m
:=
\bilin{\mathbf{f}_m,\mathbf{e}_m}.
\label{eq:sm_definition}
\end{equation}
Setting $n=m$ in Eq.~\eqref{eq:bridge} gives
\begin{equation}
s_m
=
\omega\mu
\int_\Omega
\left(
\mathbf{e}_m^-\times\mathbf{h}_m
-
\mathbf{e}_m\times\mathbf{h}_m^-
\right)
\cdot\hat{\mathbf{z}}\,d\Omega.
\label{eq:sm_bridge}
\end{equation}
Again, only the transverse field components contribute to the longitudinal component of the cross products. Since
\begin{equation}
\mathbf{e}_{m,t}^-=\mathbf{e}_{m,t},
\qquad
\mathbf{h}_{m,t}^-=-\mathbf{h}_{m,t},
\end{equation}
Eq.~\eqref{eq:sm_bridge} becomes
\begin{align}
s_m
&=
\omega\mu
\int_\Omega
\left[
\mathbf{e}_{m,t}\times\mathbf{h}_{m,t}
+
\mathbf{e}_{m,t}\times\mathbf{h}_{m,t}
\right]
\cdot\hat{\mathbf{z}}\,d\Omega\\
&=
2\omega\mu
\int_\Omega
\left(
\mathbf{e}_{m,t}\times\mathbf{h}_{m,t}
\right)
\cdot\hat{\mathbf{z}}\,d\Omega.
\label{eq:sm_final}
\end{align}
The self-pairing is therefore proportional to the unconjugated longitudinal electromagnetic flux of the mode. Since no complex conjugation appears between the electric and magnetic fields, this quantity is a modal normalization factor and should not be confused with the physical time-averaged power flux.

The same normalization factor has a useful interpretation directly within the quadratic eigenvalue problem. From the definition of the pencil pairing,
\begin{equation}
s_m
=
\langle\mathbf{f}_m,\B\mathbf{e}_m\rangle
+
2\beta_m
\langle\mathbf{f}_m,\C\mathbf{e}_m\rangle.
\end{equation}
On the other hand, differentiating the quadratic pencil with respect to $\beta$ gives
\begin{equation}
\boldsymbol{L}'(\beta)
=
\B+2\beta \C.
\end{equation}
The mode normalization can therefore be written compactly as
\begin{equation}
s_m
=
\langle
\mathbf{f}_m,
\boldsymbol{L}'(\beta_m)\mathbf{e}_m
\rangle.
\label{eq:sm_operator}
\end{equation}
This is the standard normalization factor associated with a nonlinear eigenvalue problem and also provides a direct connection to the behavior at an exceptional point.

To see this connection explicitly, suppose that $\beta_m$ is a defective eigenvalue. The corresponding eigenmode $\mathbf{e}_m$ then admits an associated vector $\mathbf{e}_m^{(1)}$ satisfying the Jordan-chain relation
\begin{equation}
\boldsymbol{L}(\beta_m)\mathbf{e}_m^{(1)}
+
\boldsymbol{L}'(\beta_m)\mathbf{e}_m
=
0.
\label{eq:jordan_chain}
\end{equation}
Taking the inner product with the corresponding left eigenmode $\mathbf{f}_m$ gives
\begin{equation}
\langle
\mathbf{f}_m,
\boldsymbol{L}(\beta_m)\mathbf{e}_m^{(1)}
\rangle
+
\langle
\mathbf{f}_m,
\boldsymbol{L}'(\beta_m)\mathbf{e}_m
\rangle
=
0.
\end{equation}
Using the adjoint relation, the first term can be written as
\begin{equation}
\langle
\mathbf{f}_m,
\boldsymbol{L}(\beta_m)\mathbf{e}_m^{(1)}
\rangle
=
\langle
\boldsymbol{L}^{\dagger}(\beta_m^*)\mathbf{f}_m,
\mathbf{e}_m^{(1)}
\rangle
=
0,
\end{equation}
because $\mathbf{f}_m$ satisfies the left eigenmode equation. The second term is $s_m$ by Eq.~\eqref{eq:sm_operator}. It therefore follows that
\begin{equation}
s_m=0.
\end{equation}
Thus, at a defective eigenvalue the mode becomes self-orthogonal with respect to the pencil pairing. Modal projection formulas involving $1/s_m$ consequently become singular, and the ordinary biorthogonal expansion must be replaced by a Jordan-chain treatment. The condition $s_m=0$ refers to the unconjugated modal flux in Eq.~\eqref{eq:sm_final} and does not imply that the physical time-averaged power carried by the mode vanishes.

Finally, we consider the lossless limit and show explicitly how the familiar
Hermitian electromagnetic relations are recovered from the general framework
developed above. Let the permittivity be real,
\begin{equation}
\epsilon(x,y)\in\mathbb{R}.
\end{equation}
In this case, the operator $\A$ becomes self-adjoint,
\begin{equation}
\A^\dagger=\A.
\end{equation}
The remaining operators already satisfy
\begin{equation}
\B^\dagger=\B,
\qquad
\C^\dagger=\C,
\end{equation}
independently of whether the permittivity is real or complex, as shown in
Appendix~\ref{app:opid}. Therefore, in the lossless case the entire quadratic
pencil is self-adjoint for real $\beta$:
\begin{equation}
\boldsymbol{L}^\dagger(\beta)
=
\A^\dagger+\beta \B^\dagger+\beta^2 \C^\dagger
=
\A+\beta \B+\beta^2 \C
=
\boldsymbol{L}(\beta).
\label{eq:lossless_selfadjoint}
\end{equation}

Consider now a guided mode with real propagation constant $\beta_m$. Its right
eigenmode satisfies
\begin{equation}
\boldsymbol{L}(\beta_m)\mathbf{e}_m=0,
\label{eq:lossless_right}
\end{equation}
whereas the corresponding left eigenmode satisfies
\begin{equation}
\boldsymbol{L}^\dagger(\beta_m^*)\mathbf{f}_m=0.
\label{eq:lossless_left}
\end{equation}
Since $\beta_m^*=\beta_m$ and
$\boldsymbol{L}^\dagger(\beta_m)=\boldsymbol{L}(\beta_m)$, the left eigenmode
obeys exactly the same equation as the right eigenmode:
\begin{equation}
\boldsymbol{L}(\beta_m)\mathbf{f}_m=0.
\end{equation}
For a simple eigenvalue, the corresponding eigenspace is one dimensional, and
therefore
\begin{equation}
\mathbf{f}_m=\alpha_m\mathbf{e}_m,
\label{eq:f_alpha_e}
\end{equation}
where $\alpha_m$ is a nonzero complex constant. Because the overall phase and
normalization of an eigenmode are arbitrary, we may choose them such that
$\alpha_m=1$. With this convention,
\begin{equation}
\boxed{\mathbf{f}_m=\mathbf{e}_m}.
\label{eq:f_equals_e_lossless}
\end{equation}
Thus, in the lossless limit, the distinction between the left and right
eigenmodes disappears, as expected for a self-adjoint eigenvalue problem.

This result can now be combined with the general relation between the left
eigenmode and the backward-propagating partner derived above,
\begin{equation}
\mathbf{f}_m=(\mathbf{e}_m^-)^*.
\label{eq:f_backward_again}
\end{equation}
Using Eq.~\eqref{eq:f_equals_e_lossless}, we obtain
\begin{equation}
\mathbf{e}_m=(\mathbf{e}_m^-)^*,
\end{equation}
or equivalently,
\begin{equation}
\boxed{\mathbf{e}_m^-=\mathbf{e}_m^*.}
\label{eq:backward_conjugate_lossless}
\end{equation}

The componentwise consequences of this relation are particularly transparent.
We have already shown that the forward- and backward-propagating electric-field
profiles are related by
\begin{equation}
\mathbf{e}_m
=
\mathbf{e}_{m,t}
+
\hat{\mathbf{z}}e_{m,z},
\qquad
\mathbf{e}_m^-
=
\mathbf{e}_{m,t}
-
\hat{\mathbf{z}}e_{m,z}.
\label{eq:forward_backward_components_lossless}
\end{equation}
On the other hand, Eq.~\eqref{eq:backward_conjugate_lossless} gives
\begin{equation}
\mathbf{e}_m^-
=
\mathbf{e}_{m,t}^*
+
\hat{\mathbf{z}}e_{m,z}^*.
\end{equation}
Equating the transverse and longitudinal components of these two expressions
separately yields
\begin{equation}
\mathbf{e}_{m,t}
=
\mathbf{e}_{m,t}^*,
\qquad
-e_{m,z}
=
e_{m,z}^*.
\end{equation}
Hence,
\begin{equation}
\boxed{
\mathbf{e}_{m,t}\in\mathbb{R},
\qquad
e_{m,z}\in i\mathbb{R}.
}
\label{eq:lossless_phase_parity}
\end{equation}
Thus, after fixing the arbitrary global phase of the mode, the transverse
electric-field components can be chosen real while the longitudinal component
is purely imaginary. The familiar $90^\circ$ relative phase between the
transverse and longitudinal electric-field components therefore emerges
directly from the relation between the left eigenmode and the
backward-propagating partner in the self-adjoint limit.

The corresponding relation for the magnetic field follows from Faraday's law.
As shown in Appendix~\ref{app:magnetic}, in the same phase convention the
forward- and backward-propagating magnetic-field profiles satisfy
\begin{equation}
\boxed{
\mathbf{h}_m^-=-\mathbf{h}_m^*.
}
\label{eq:backward_H_conjugate_lossless}
\end{equation}
We can therefore return to the unconjugated Lorentz modal orthogonality
relation obtained above,
\begin{equation}
\int_\Omega
\left(
\mathbf{e}_m^-\times\mathbf{h}_n
-
\mathbf{e}_n\times\mathbf{h}_m^-
\right)
\cdot\hat{\mathbf{z}}\,d\Omega
=0,
\qquad
\beta_n\neq\beta_m.
\end{equation}
Substituting
$\mathbf{e}_m^-=\mathbf{e}_m^*$ and
$\mathbf{h}_m^-=-\mathbf{h}_m^*$ gives
\begin{equation}
\int_\Omega
\left(
\mathbf{e}_m^*\times\mathbf{h}_n
+
\mathbf{e}_n\times\mathbf{h}_m^*
\right)
\cdot\hat{\mathbf{z}}\,d\Omega
=0,
\qquad
\beta_n\neq\beta_m.
\label{eq:conjugated_lorentz}
\end{equation}
We therefore recover the familiar conjugated electromagnetic modal
orthogonality relation in the lossless limit. This provides a direct
consistency check of the framework: the adjoint-based biorthogonal formulation
for non-Hermitian Maxwell eigenmodes reduces continuously to the conventional
Hermitian modal structure when gain and loss are removed.

\section{Conclusion}
\label{sec:discussion}

In this work, we revisited the notions of left and right eigenstates for non-Hermitian electromagnetic systems directly within the full vectorial Maxwell framework. Starting from the electric-field wave equation for guided modes, we formulated the problem as a quadratic eigenvalue problem in the propagation constant and constructed the corresponding adjoint problem. This allowed us to identify the left eigenmodes and establish the associated biorthogonality relation without relying on scalar, paraxial, or coupled-mode descriptions.

A central result of our analysis is the explicit connection between this operator-theoretic biorthogonality and the conventional modal orthogonality relation obtained from unconjugated Lorentz reciprocity. We showed that the left eigenmode can be identified with the complex conjugate of the corresponding backward-propagating electric-field mode and that, through Maxwell's equations, the electric-field-only biorthogonal pairing is exactly related to the mixed electric--magnetic field expression appearing in Lorentz reciprocity. The two approaches therefore provide equivalent descriptions of the same underlying electromagnetic modal structure.

The framework also provides a direct interpretation of the modal normalization and its behavior at exceptional points. In particular, the self-pairing that normalizes the left and right eigenmodes is directly related to the unconjugated longitudinal electromagnetic flux and naturally vanishes when an eigenmode becomes defective, recovering the familiar self-orthogonality associated with exceptional points. In the lossless limit, the formulation consistently reduces to the conventional Hermitian description, including the familiar relations between forward- and backward-propagating fields and the standard conjugated electromagnetic modal orthogonality relation.

Taken together, these results provide a unified operator-theoretic interpretation of electromagnetic biorthogonality and Lorentz reciprocity for full-vectorial non-Hermitian guided waves. By establishing a direct bridge between the left/right eigenstate language widely used in non-Hermitian physics and the reciprocity-based modal framework traditionally used in electromagnetics, this work provides a consistent foundation for modal expansions and related analyses of vectorial non-Hermitian photonic systems.

\begin{acknowledgments}
R.E. acknowledges support from the AFOSR Multidisciplinary University Research Initiative Award on Programmable Systems with Non-Hermitian Quantum Dynamics (Grant No.FA9550-21-1-0202). R.E. also acknowledges support from Army Research Office (W911NF-23-1-0312).
\end{acknowledgments}

\appendix

\section{Conjugated and Unconjugated Lorentz Reciprocity}
\label{app:lorentz}

Let $(\mathbf{E}_1,\mathbf{H}_1)$ and $(\mathbf{E}_2,\mathbf{H}_2)$ be two
source-free time-harmonic fields at the same frequency $\omega$ in a
nonmagnetic medium with relative permittivity $\eps(x,y)$, possibly
complex.  With the $e^{-i\omega t}$ convention, Maxwell's curl equations
read $\nabla\times\mathbf{E}=i\omega\mu_0\mathbf{H}$ and
$\nabla\times\mathbf{H}=-i\omega\eps_0\eps\,\mathbf{E}$.

\emph{Unconjugated form.}\;
Using $\nabla\cdot(\mathbf{U}\times\mathbf{V})
=\mathbf{V}\cdot(\nabla\times\mathbf{U})
-\mathbf{U}\cdot(\nabla\times\mathbf{V})$,
\begin{multline}
\nabla\cdot(\mathbf{E}_1\times\mathbf{H}_2-\mathbf{E}_2\times\mathbf{H}_1)
= i\omega\mu_0(\mathbf{H}_1\cdot\mathbf{H}_2
-\mathbf{H}_2\cdot\mathbf{H}_1)\\
+ i\omega\eps_0\eps\,(\mathbf{E}_1\cdot\mathbf{E}_2
-\mathbf{E}_2\cdot\mathbf{E}_1)=0,
\end{multline}
which holds for \emph{arbitrary complex} $\eps$ (only symmetry of the
constitutive tensors is required).  For two modes
$\mathbf{E}_a=\mathbf{e}_a(x,y)e^{i\beta_a z}$ with signed propagation
constants, integrating over the cross-section $\Omega$ and using
$z$-invariance together with the decay of guided fields gives
\begin{equation}
(\beta_a+\beta_b)\int_\Omega
(\mathbf{e}_a\times\mathbf{h}_b-\mathbf{e}_b\times\mathbf{h}_a)
\cdot\hat{\mathbf{z}}\,d\Omega=0,
\end{equation}
so the profile integral vanishes whenever $\beta_a+\beta_b\neq0$.  The pairing can be non-zero only between a mode and
its backward partner ($\beta_b=-\beta_a$), which is the case evaluated in Eq.~\eqref{eq:sm_bridge}.

\emph{Conjugated form.}\;
Repeating the computation for
$\mathbf{E}_1\times\mathbf{H}_2^{*}+\mathbf{E}_2^{*}\times\mathbf{H}_1$
and using $\nabla\times\mathbf{H}_2^{*}=+i\omega\eps_0\eps^{*}\mathbf{E}_2^{*}$
yields
\begin{equation}
\nabla\cdot(\mathbf{E}_1\times\mathbf{H}_2^{*}
+\mathbf{E}_2^{*}\times\mathbf{H}_1)
= i\omega\eps_0(\eps-\eps^{*})\,\mathbf{E}_1\cdot\mathbf{E}_2^{*},
\end{equation}
which vanishes only for real $\eps$.  In that lossless case the same
$z$-invariance argument gives
$(\beta_a-\beta_b^{*})\int_\Omega
(\mathbf{e}_a\times\mathbf{h}_b^{*}
+\mathbf{e}_b^{*}\times\mathbf{h}_a)\cdot\hat{\mathbf{z}}\,d\Omega=0$,
i.e., power orthogonality for real, distinct propagation constants.  This
makes explicit why the conjugated relation is tied to Hermiticity, while
the unconjugated relation survives in the presence of gain and loss.

\section{Pencil Equations}
\label{app:pencilEquations}
We first consider the action of $\mathcal{A}$ defined in Eq~\eqref{eq:Aoperator}. By 
substituting $\mathbf{e}=\mathbf{e}_t+\hat{\mathbf{z}}e_z$ gives
\begin{multline}
\mathcal{A}\mathbf{e}
=
\nabla_t\times(\nabla_t\times\mathbf{e}_t)
+\nabla_t\times[\nabla_t\times(\hat{\mathbf{z}}e_z)]
\\-k_0^2\epsilon\mathbf{e}_t
-k_0^2\epsilon\hat{\mathbf{z}}e_z.
\end{multline}

Since $\nabla_t$ contains only derivatives with respect to $x$ and $y$,
\begin{equation}
\nabla_t\times(\hat{\mathbf{z}}e_z)
=
\nabla_t e_z\times\hat{\mathbf{z}},
\end{equation}
and a second transverse curl gives
\begin{equation}
\nabla_t\times[\nabla_t\times(\hat{\mathbf{z}}e_z)]
=
-\hat{\mathbf{z}}\nabla_t^2 e_z.
\end{equation}
Hence,
\begin{align}
\mathcal{A}\mathbf{e}
&=
\left[
\nabla_t\times(\nabla_t\times\mathbf{e}_t)
-k_0^2\epsilon\mathbf{e}_t
\right]
-\hat{\mathbf{z}}
\left(
\nabla_t^2+k_0^2\epsilon
\right)e_z.
\end{align}
Defining
\begin{equation}
\mathcal{T}\equiv\nabla_t\times\nabla_t\times-k_0^2\epsilon,
\qquad
\mathcal{S}\equiv\nabla_t^2+k_0^2\epsilon,
\end{equation}
we obtain
\begin{equation}
\mathcal{A}\mathbf{e}
=
\mathcal{T}\mathbf{e}_t
-\hat{\mathbf{z}}\mathcal{S}e_z.
\end{equation}

We next consider the operator $\mathcal{B}$, defined in Eq.~\eqref{eq:Boperator}

To evaluate the first term, we use the vector identity
\begin{equation}
\nabla\times(\mathbf{a}\times\mathbf{b})
=
\mathbf{a}(\nabla\cdot\mathbf{b})
-\mathbf{b}(\nabla\cdot\mathbf{a})
+(\mathbf{b}\cdot\nabla)\mathbf{a}
-(\mathbf{a}\cdot\nabla)\mathbf{b}.
\end{equation}
Taking $\mathbf{a}=\hat{\mathbf{z}}$, $\mathbf{b}=\mathbf{e}$, and $\nabla=\nabla_t$, and noting that $\hat{\mathbf{z}}$ is constant and $\hat{\mathbf{z}}\cdot\nabla_t=0$, we obtain
\begin{equation}
\nabla_t\times(\hat{\mathbf{z}}\times\mathbf{e})
=
\hat{\mathbf{z}}(\nabla_t\cdot\mathbf{e})
=
\hat{\mathbf{z}}(\nabla_t\cdot\mathbf{e}_t).
\end{equation}

For the second term, we first decompose the field as $\mathbf{e}=\mathbf{e}_t+\hat{\mathbf{z}}e_z$, which gives
\begin{equation}
\nabla_t\times\mathbf{e}
=
\nabla_t\times\mathbf{e}_t
+
\nabla_t\times(\hat{\mathbf{z}}e_z).
\end{equation}
Since $\mathbf{e}_t$ and $\nabla_t$ are both transverse, $\nabla_t\times\mathbf{e}_t$ is directed along $\hat{\mathbf{z}}$, and hence
\begin{equation}
\hat{\mathbf{z}}\times(\nabla_t\times\mathbf{e}_t)=0.
\end{equation}
Furthermore,
\begin{equation}
\nabla_t\times(\hat{\mathbf{z}}e_z)
=
\nabla_t e_z\times\hat{\mathbf{z}}.
\end{equation}
It follows that
\begin{equation}
\hat{\mathbf{z}}\times(\nabla_t\times\mathbf{e})
=
\hat{\mathbf{z}}\times(\nabla_t e_z\times\hat{\mathbf{z}}).
\end{equation}
Using the vector triple-product identity
\begin{equation}
\mathbf{a}\times(\mathbf{b}\times\mathbf{c})
=
\mathbf{b}(\mathbf{a}\cdot\mathbf{c})
-
\mathbf{c}(\mathbf{a}\cdot\mathbf{b}),
\end{equation}
together with $\hat{\mathbf{z}}\cdot\nabla_t e_z=0$, we find
\begin{equation}
\hat{\mathbf{z}}\times(\nabla_t\times\mathbf{e})
=
\nabla_t e_z.
\end{equation}
Combining the two contributions therefore gives
\begin{equation}\label{eq:Bexplicit}
\mathcal{B}\mathbf{e}
=
i\left[
\nabla_t e_z
+
\hat{\mathbf{z}}(\nabla_t\cdot\mathbf{e}_t)
\right].
\end{equation}

Finally,
\begin{equation}\label{eq:Cexplicit}
\mathcal{C}\mathbf{e}
\equiv
-\hat{\mathbf{z}}\times(\hat{\mathbf{z}}\times\mathbf{e}).
\end{equation}
Since
\begin{equation}
\hat{\mathbf{z}}\times
(\hat{\mathbf{z}}\times\mathbf{e})
=
\hat{\mathbf{z}}e_z-\mathbf{e}
=
-\mathbf{e}_t,
\end{equation}
we immediately obtain
\begin{equation}
\mathcal{C}\mathbf{e}=\mathbf{e}_t.
\end{equation}

\section{Block GEVP Derivation}\label{app:gevp}
 Inserting the explicit operator forms into the pencil
$\boldsymbol{L}(\beta)\mathbf{e}=0$ and separating transverse
($\hat{\mathbf{z}}$-free) and longitudinal ($\hat{\mathbf{z}}$) parts yields:
\begin{equation}
\mathcal{T}\mathbf{e}_t + i\beta\nabla_t e_z + \beta^2\mathbf{e}_t = 0,
\label{eq:trans_coupled}
  \end{equation}
\begin{equation}
\mathcal{S}e_z - i\beta(\nabla_t\cdot\mathbf{e}_t) = 0.
\label{eq:long_coupled}
\end{equation}
The substitution $\tilde{e}_z=i\beta e_z$ converts both equations into
a form quadratic in $\beta$ only through the common factor $\beta^2$:
\begin{equation}
\mathcal{T}\mathbf{e}_t + \nabla_t\tilde{e}_z = -\beta^2\mathbf{e}_t,
  \end{equation}
  \begin{equation}
\mathcal{S}\tilde{e}_z = -\beta^2(\nabla_t\cdot\mathbf{e}_t).
\end{equation}
Setting $\lambda=\beta^2$ and collecting into block matrix form gives the
block GEVP~\eqref{eq:block_gevp} of the main text. Since $\beta$ enters the GEVP only through
$\lambda=\beta^2$, each eigenpair $(\lambda;\mathbf{e}_t,\tilde{e}_z)$
generates two physical modes $\beta=\pm\sqrt{\lambda}$.  Undoing the
substitution, $e_z=\tilde{e}_z/(i\beta)$, shows that both roots share the
same transverse profile $\mathbf{e}_t$ while $e_z$ changes sign with
$\beta$, reproducing the $\pm$ pairing of Eq.~\eqref{eq:pm_modes}.

\section{Proof of Operator Identities}
\label{app:opid}
We prove the three identities used to identify the adjoint mode.

\noindent\textit{$(\mathcal{A}^\dagger)^* = \mathcal{A}$.}\;
Evaluating $\langle\mathbf{f},\mathcal{A}\mathbf{e}\rangle$ and applying the
divergence identity
$\nabla\cdot(\mathbf{U}\times\mathbf{V})
=\mathbf{V}\cdot(\nabla\times\mathbf{U})-\mathbf{U}\cdot(\nabla\times\mathbf{V})$
twice (with boundary terms vanishing under guided-mode decay) transfers
both curls onto $\mathbf{f}^*$:
\begin{equation}
\int_\Omega\mathbf{f}^*\cdot(\nabla_t\times\nabla_t\times\mathbf{e})\,d\Omega
=\int_\Omega(\nabla_t\times\nabla_t\times\mathbf{f})^*\cdot\mathbf{e}\,d\Omega,
\end{equation}
using the fact that $\nabla_t$ (with real Cartesian coefficients) commutes
with complex conjugation.  The multiplication term gives
$\int_\Omega\mathbf{f}^*\cdot(-k_0^2 \eps\mathbf{e})\,d\Omega
=\int_\Omega(-k_0^2 \eps^{*}\mathbf{f})^*\cdot\mathbf{e}\,d\Omega$
for $k_0\in\mathbb{R}$.
Hence $\A^\dagger=\nabla_t\times\nabla_t\times-k_0^2\eps^{*}$,
and taking the complex conjugate:
$(\mathcal{A}^\dagger)^*=\nabla_t\times\nabla_t\times-k_0^2 \eps=\mathcal{A}$.

\noindent\textit{$(\mathcal{B}^\dagger)^* = -\mathcal{B}$}\;
Using the explicit form $\mathcal{B}\mathbf{e}=i[\nabla_t e_z+\hat{\mathbf{z}}(\nabla_t\cdot\mathbf{e}_t)]$,
one evaluates $\langle\mathbf{f},\B\mathbf{e}\rangle$ and integrates
by parts (boundary terms vanish):
\begin{align}
\langle\mathbf{f},\mathcal{B}\mathbf{e}\rangle
&= i\int_\Omega\bigl[\mathbf{f}_t^*\cdot\nabla_t e_z
   +f_z^*(\nabla_t\cdot\mathbf{e}_t)\bigr]d\Omega \notag\\
&= -i\int_\Omega\bigl[(\nabla_t\cdot\mathbf{f}_t^*)e_z
   +(\nabla_t f_z^*)\cdot\mathbf{e}_t\bigr]d\Omega.
\end{align}
An identical calculation of $\langle \B\mathbf{f},\mathbf{e}\rangle$ gives
the same integral, so $\mathcal{B}^\dagger=\mathcal{B}$.  Since $\mathcal{B}$ carries an explicit factor
of $i$: $(\mathcal{B}^\dagger)^*=\mathcal{B}^*=-\mathcal{B}$.

\noindent\textit{$(\mathcal{C}^\dagger)^* = \mathcal{C}$}\;
Since $\mathcal{C}\mathbf{e}=\mathbf{e}_t$ is a real projection,
$\langle\mathbf{f},\mathcal{C}\mathbf{e}\rangle
=\int_\Omega\mathbf{f}_t^*\cdot\mathbf{e}_t\,d\Omega
=\langle \mathcal{C}\mathbf{f},\mathbf{e}\rangle$,
so $\mathcal{C}^\dagger=\mathcal{C}$ and $(\mathcal{C}^\dagger)^*=\mathcal{C}$.

\section{Derivation of the Bridge Relation and Mode Norm}
\label{app:bridge}

\subsection{Bridge relation}
\label{app:bridgeproof}

Using $(\mathbf{A}\times\mathbf{B})\cdot\hat{\mathbf{z}}
=-\mathbf{A}\cdot(\hat{\mathbf{z}}\times\mathbf{B})$ and retaining only
transverse contributions:
\begin{equation}
(\mathbf{e}^-_m\times\mathbf{h}_n
-\mathbf{e}_n\times\mathbf{h}^-_m)\cdot\hat{\mathbf{z}}
=-\mathbf{e}^-_{m,t}\cdot(\hat{\mathbf{z}}\times\mathbf{h}_{n,t})
+\mathbf{e}_{n,t}\cdot(\hat{\mathbf{z}}\times\mathbf{h}^-_{m,t}).
\label{eq:fluxform}
\end{equation}
From Eq.~\eqref{eq:Ht_fwd}, the transverse magnetic fields expressed via
Faraday's law are:
\begin{equation}
\hat{\mathbf{z}}\times\mathbf{h}_{n,t}
  =\frac{1}{i\omega\mu}(\nabla_t e_{n,z}-i\beta_n\mathbf{e}_{n,t}),
  \label{eq:hn}
  \end{equation}
  \begin{equation}
\hat{\mathbf{z}}\times\mathbf{h}^-_{m,t}
  =\frac{1}{i\omega\mu}(\nabla_t e^-_{m,z}+i\beta_m\mathbf{e}^-_{m,t}).
  \label{eq:hm}
\end{equation}
Substituting~\eqref{eq:hn}--\eqref{eq:hm} into~\eqref{eq:fluxform},
using $\mathbf{e}^-_m=\mathbf{f}_m^*$ (so
$\mathbf{e}^-_{m,t}=\mathbf{f}_{m,t}^*$ and
$e^-_{m,z}=-e_{m,z}=f_{m,z}^{*}$, hence
$\nabla_t e^-_{m,z}=\nabla_t f_{m,z}^{*}$), and integrating
over $\Omega$:
\begin{multline}
\int_\Omega(\mathbf{e}^-_m\times\mathbf{h}_n
-\mathbf{e}_n\times\mathbf{h}^-_m)\cdot\hat{\mathbf{z}}\,d\Omega
= \frac{1}{i\omega\mu}\int_\Omega\left[
-\mathbf{f}_{m,t}^*\cdot\nabla_t e_{n,z} \right.\ \\ \left.
+\mathbf{e}_{n,t}\cdot\nabla_t f_{m,z}^*
+i(\beta_n+\beta_m)\mathbf{f}_{m,t}^*\cdot\mathbf{e}_{n,t}
\right]d\Omega.
\end{multline}
Integrating by parts on the $\nabla_t f_{m,z}^*$ term (boundary terms
vanish):
$\int\mathbf{e}_{n,t}\cdot\nabla_t f_{m,z}^*
=-\int f_{m,z}^*(\nabla_t\cdot\mathbf{e}_{n,t})\,d\Omega$.
The right-hand side becomes $i$ times the explicit form of the bilinear
pairing $\bilin{\mathbf{f}_m,\mathbf{e}_n}$
from Eqs.~\eqref{eq:Bexplicit},~\eqref{eq:Cexplicit},
and~\eqref{eq:bilinearF}, yielding the bridge
relation~\eqref{eq:bridge}.

\subsection{Mode norm}
Setting $n=m$ in the bridge relation and using $\mathbf{e}^-_{m,t}=\mathbf{e}_{m,t}$
and $\mathbf{h}^-_{m,t}=-\mathbf{h}_{m,t}$:
\begin{equation}
(\mathbf{e}^-_m\times\mathbf{h}_m)\cdot\hat{\mathbf{z}}
  =(\mathbf{e}_{m,t}\times\mathbf{h}_{m,t})\cdot\hat{\mathbf{z}},
  \end{equation}
  \begin{equation}
-(\mathbf{e}_m\times\mathbf{h}^-_m)\cdot\hat{\mathbf{z}}
  =(\mathbf{e}_{m,t}\times\mathbf{h}_{m,t})\cdot\hat{\mathbf{z}}.
\end{equation}
Summing and inserting into Eq.~\eqref{eq:bridge} gives
$s_m=2\omega\mu\int_\Omega(\mathbf{e}_{m,t}\times\mathbf{h}_{m,t})
\cdot\hat{\mathbf{z}}\,d\Omega$.


\section{Backward Magnetic Field from Faraday's Law}
\label{app:magnetic}

Applying the modal ansatz $\mathbf{H}=\mathbf{h}(x,y)e^{i\beta z}$ to
Faraday's law $\nabla\times\mathbf{E}=i\omega\mu\mathbf{H}$
and substituting $\nabla\to\nabla_t+i\beta\hat{\mathbf{z}}$ yields,
upon separating transverse and longitudinal components:
\begin{equation}
i\omega\mu\,h_z\hat{\mathbf{z}} = \nabla_t\times\mathbf{e}_t,
\label{eq:Hz_fwd}
\end{equation}
\begin{equation}
i\omega\mu\,\mathbf{h}_t = \nabla_t e_z\times\hat{\mathbf{z}}
                            + i\beta\,\hat{\mathbf{z}}\times\mathbf{e}_t.
\label{eq:Ht_fwd}
\end{equation}
For the backward mode $\mathbf{e}^-$ with $\mathbf{e}^-_t=\mathbf{e}_t$,
$e^-_z=-e_z$, and $\beta\to-\beta$:
Eq.~\eqref{eq:Hz_fwd} is unchanged since its right-hand side depends only
on $\mathbf{e}_t$, giving $h^-_z=h_z$.
Substituting into Eq.~\eqref{eq:Ht_fwd} and factoring:
$i\omega\mu\,\mathbf{h}^-_t
=-(\nabla_t e_z\times\hat{\mathbf{z}}+i\beta\,\hat{\mathbf{z}}\times\mathbf{e}_t)$,
so $\mathbf{h}^-_t=-\mathbf{h}_t$. From Faraday's law, and a mode $m$,
$\mathbf{e}_{m,t}$ real and $e_{m,z}$ purely imaginary,
Eqs.~\eqref{eq:Hz_fwd}--\eqref{eq:Ht_fwd} make $\mathbf{h}_{m,t}$ real and
$h_{m,z}$ purely imaginary, so both components of
$\mathbf{h}^-_m$ obey the same relation, and we get $\mathbf{h}_m^-=-\mathbf{h}_m^*$.

\bibliographystyle{unsrt}

\bibliography{references}

@article{curtright2007biorthogonal,
  author  = {Curtright, T. and Mezincescu, L.},
  title   = {Biorthogonal quantum systems},
  journal = {J. Math. Phys.},
  volume  = {48},
  number  = {9},
  pages   = {092106},
  year    = {2007},
  doi     = {10.1063/1.2196243}
}

@article{driscoll2009longitudinal,
  author  = {Driscoll, J. B. and Liu, X. and Yasseri, S.
             and Hsieh, I. and Dadap, J. I.
             and Osgood, Jr., Richard M.},
  title   = {Large longitudinal electric fields ({$E_z$})
             in silicon nanowire waveguides},
  journal = {Opt. Express},
  volume  = {17},
  number  = {4},
  pages   = {2797--2804},
  year    = {2009},
  doi     = {10.1364/OE.17.002797}
}

@article{afshar2009vectorial,
  author  = {{Afshar V.}, S. and Monro, T. M.},
  title   = {A full vectorial model for pulse propagation in emerging
             waveguides with subwavelength structures part {I}:
             {K}err nonlinearity},
  journal = {Opt. Express},
  volume  = {17},
  number  = {4},
  pages   = {2298--2318},
  year    = {2009},
  doi     = {10.1364/OE.17.002298}
}

@article{truong2020dqnm,
  author  = {Truong, M. D. and Nicolet, A.
             and Dem{\'e}sy, G. and Zolla, F.},
  title   = {Continuous family of exact Dispersive Quasi-Normal
             Modal ({DQNM}) expansions for dispersive photonic structures},
  journal = {Opt. Express},
  volume  = {28},
  number  = {20},
  pages   = {29016--29032},
  year    = {2020},
  doi     = {10.1364/OE.401742}
}

@article{turner2009vectorial,
  author  = {Turner, M. D. and Monro, T. M. and {Afshar V.}, S.},
  title   = {A full vectorial model for pulse propagation in emerging
             waveguides with subwavelength structures part {II}:
             {S}timulated {R}aman scattering},
  journal = {Opt. Express},
  volume  = {17},
  number  = {14},
  pages   = {11565--11581},
  year    = {2009},
  doi     = {10.1364/OE.17.011565}
}

@article{elganainy2018nhpt,
  author  = {El-Ganainy, R. and Makris, K. G. and Khajavikhan, M.
             and Musslimani, Z. H. and Rotter, S.
             and Christodoulides, D. N.},
  title   = {Non-{H}ermitian physics and {PT} symmetry},
  journal = {Nat. Phys.},
  volume  = {14},
  number  = {1},
  pages   = {11--19},
  year    = {2018},
  doi     = {10.1038/nphys4323}
}

@article{feng2017nhphotonics,
  author  = {Feng, L. and El-Ganainy, R. and Ge, L.},
  title   = {Non-{H}ermitian photonics based on parity--time symmetry},
  journal = {Nat. Photonics},
  volume  = {11},
  number  = {12},
  pages   = {752--762},
  year    = {2017},
  doi     = {10.1038/s41566-017-0031-1}
}

@article{miri2019eps,
  author  = {Miri, M.-A. and Al{\`u}, A.},
  title   = {Exceptional points in optics and photonics},
  journal = {Science},
  volume  = {363},
  number  = {6422},
  pages   = {eaar7709},
  year    = {2019},
  doi     = {10.1126/science.aar7709}
}

@article{ozdemir2019ptep,
  author  = {{\"O}zdemir, {\c S}. K. and Rotter, S. and Nori, F. and Yang, L.},
  title   = {Parity--time symmetry and exceptional points in photonics},
  journal = {Nat. Mater.},
  volume  = {18},
  number  = {8},
  pages   = {783--798},
  year    = {2019},
  doi     = {10.1038/s41563-019-0304-9}
}

@article{bender1998real,
  author  = {Bender, C. M. and Boettcher, S.},
  title   = {Real spectra in non-{H}ermitian {H}amiltonians having {PT}
             symmetry},
  journal = {Phys. Rev. Lett.},
  volume  = {80},
  number  = {24},
  pages   = {5243--5246},
  year    = {1998},
  doi     = {10.1103/PhysRevLett.80.5243}
}

@article{bender2007making,
  author  = {Bender, C. M.},
  title   = {Making sense of non-{H}ermitian {H}amiltonians},
  journal = {Rep. Prog. Phys.},
  volume  = {70},
  number  = {6},
  pages   = {947--1018},
  year    = {2007},
  doi     = {10.1088/0034-4885/70/6/R03}
}

@article{mostafazadeh2002pseudo,
  author  = {Mostafazadeh, A.},
  title   = {Pseudo-{H}ermiticity versus {PT} symmetry: {T}he necessary
             condition for the reality of the spectrum of a non-{H}ermitian
             {H}amiltonian},
  journal = {J. Math. Phys.},
  volume  = {43},
  number  = {1},
  pages   = {205--214},
  year    = {2002},
  doi     = {10.1063/1.1418246}
}

@article{elganainy2007ptcmt,
  author  = {El-Ganainy, R. and Makris, K. G. and Christodoulides, D. N.
             and Musslimani, Z. H.},
  title   = {Theory of coupled optical {PT}-symmetric structures},
  journal = {Opt. Lett.},
  volume  = {32},
  number  = {17},
  pages   = {2632--2634},
  year    = {2007},
  doi     = {10.1364/OL.32.002632}
}

@article{makris2008beam,
  author  = {Makris, K. G. and El-Ganainy, R. and Christodoulides, D. N.
             and Musslimani, Z. H.},
  title   = {Beam dynamics in {PT}-symmetric optical lattices},
  journal = {Phys. Rev. Lett.},
  volume  = {100},
  number  = {10},
  pages   = {103904},
  year    = {2008},
  doi     = {10.1103/PhysRevLett.100.103904}
}

@article{guo2009observation,
  author  = {Guo, A. and Salamo, G. J. and Duchesne, D. and Morandotti, R.
             and Volatier-Ravat, M. and Aimez, V. and Siviloglou, G. A.
             and Christodoulides, D. N.},
  title   = {Observation of {PT}-symmetry breaking in complex optical
             potentials},
  journal = {Phys. Rev. Lett.},
  volume  = {103},
  number  = {9},
  pages   = {093902},
  year    = {2009},
  doi     = {10.1103/PhysRevLett.103.093902}
}

@article{peng2014wgm,
  author  = {Peng, B. and {\"O}zdemir, {\c S}. K. and Lei, F. and Monifi, F.
             and Gianfreda, M. and Long, G. L. and Fan, S. and Nori, F.
             and Bender, C. M. and Yang, L.},
  title   = {Parity--time-symmetric whispering-gallery microcavities},
  journal = {Nat. Phys.},
  volume  = {10},
  number  = {5},
  pages   = {394--398},
  year    = {2014},
  doi     = {10.1038/nphys2927}
}

@article{chang2014isolation,
  author  = {Chang, L. and Jiang, X. and Hua, S. and Yang, C. and Wen, J.
             and Jiang, L. and Li, G. and Wang, G. and Xiao, M.},
  title   = {Parity--time symmetry and variable optical isolation in
             active--passive-coupled microresonators},
  journal = {Nat. Photonics},
  volume  = {8},
  number  = {7},
  pages   = {524--529},
  year    = {2014},
  doi     = {10.1038/nphoton.2014.133}
}

@article{feng2014singlemode,
  author  = {Feng, L. and Wong, Z. J. and Ma, R.-M. and Wang, Y. and Zhang, X.},
  title   = {Single-mode laser by parity-time symmetry breaking},
  journal = {Science},
  volume  = {346},
  number  = {6212},
  pages   = {972--975},
  year    = {2014},
  doi     = {10.1126/science.1258479}
}

@article{hodaei2014microring,
  author  = {Hodaei, H. and Miri, M.-A. and Heinrich, M.
             and Christodoulides, D. N. and Khajavikhan, M.},
  title   = {Parity-time--symmetric microring lasers},
  journal = {Science},
  volume  = {346},
  number  = {6212},
  pages   = {975--978},
  year    = {2014},
  doi     = {10.1126/science.1258480}
}

@article{bandres2018topological,
  author  = {Bandres, M. A. and Wittek, S. and Harari, G. and Parto, M.
             and Ren, J. and Segev, M. and Christodoulides, D. N.
             and Khajavikhan, M.},
  title   = {Topological insulator laser: {E}xperiments},
  journal = {Science},
  volume  = {359},
  number  = {6381},
  pages   = {eaar4005},
  year    = {2018},
  doi     = {10.1126/science.aar4005}
}

@article{wiersig2014sensing,
  author  = {Wiersig, J.},
  title   = {Enhancing the sensitivity of frequency and energy splitting
             detection by using exceptional points: {A}pplication to
             microcavity sensors for single-particle detection},
  journal = {Phys. Rev. Lett.},
  volume  = {112},
  number  = {20},
  pages   = {203901},
  year    = {2014},
  doi     = {10.1103/PhysRevLett.112.203901}
}

@article{hodaei2017higherorder,
  author  = {Hodaei, H. and Hassan, A. U. and Wittek, S.
             and Garcia-Gracia, H. and El-Ganainy, R.
             and Christodoulides, D. N. and Khajavikhan, M.},
  title   = {Enhanced sensitivity at higher-order exceptional points},
  journal = {Nature (London)},
  volume  = {548},
  number  = {7666},
  pages   = {187--191},
  year    = {2017},
  doi     = {10.1038/nature23280}
}

@article{ramezani2010unidirectional,
  author  = {Ramezani, H. and Kottos, T. and El-Ganainy, R.
             and Christodoulides, D. N.},
  title   = {Unidirectional nonlinear {PT}-symmetric optical structures},
  journal = {Phys. Rev. A},
  volume  = {82},
  number  = {4},
  pages   = {043803},
  year    = {2010},
  doi     = {10.1103/PhysRevA.82.043803}
}

@article{lumer2013nonlinear,
  author  = {Lumer, Y. and Plotnik, Y. and Rechtsman, M. C. and Segev, M.},
  title   = {Nonlinearly induced {PT} transition in photonic systems},
  journal = {Phys. Rev. Lett.},
  volume  = {111},
  number  = {26},
  pages   = {263901},
  year    = {2013},
  doi     = {10.1103/PhysRevLett.111.263901}
}

@article{petermann1979,
  author  = {Petermann, K.},
  title   = {Calculated spontaneous emission factor for
             double-heterostructure injection lasers with gain-induced
             waveguiding},
  journal = {IEEE J. Quantum Electron.},
  volume  = {15},
  number  = {7},
  pages   = {566--570},
  year    = {1979},
  doi     = {10.1109/JQE.1979.1070064}
}

@article{siegman1989,
  author  = {Siegman, A. E.},
  title   = {Excess spontaneous emission in non-{H}ermitian optical
             systems. {I}. {L}aser amplifiers},
  journal = {Phys. Rev. A},
  volume  = {39},
  number  = {3},
  pages   = {1253--1263},
  year    = {1989},
  doi     = {10.1103/PhysRevA.39.1253}
}

@article{brody2014biorthogonal,
  author  = {Brody, D. C.},
  title   = {Biorthogonal quantum mechanics},
  journal = {J. Phys. A: Math. Theor.},
  volume  = {47},
  number  = {3},
  pages   = {035305},
  year    = {2014},
  doi     = {10.1088/1751-8113/47/3/035305}
}

@book{moiseyev2011,
  author    = {Moiseyev, N.},
  title     = {Non-{H}ermitian Quantum Mechanics},
  publisher = {Cambridge University Press},
  address   = {Cambridge},
  year      = {2011},
  doi       = {10.1017/CBO9780511976186}
}

@article{bresler1958,
  author  = {Bresler, A. D. and Joshi, G. H. and Marcuvitz, N.},
  title   = {Orthogonality properties for modes in passive and active
             uniform wave guides},
  journal = {J. Appl. Phys.},
  volume  = {29},
  number  = {5},
  pages   = {794--799},
  year    = {1958}
}

@book{snyder1983,
  author    = {Snyder, A. W. and Love, J. D.},
  title     = {Optical Waveguide Theory},
  publisher = {Chapman and Hall},
  address   = {London},
  year      = {1983}
}

@book{collin1991,
  author    = {Collin, R. E.},
  title     = {Field Theory of Guided Waves},
  edition   = {2},
  publisher = {IEEE Press},
  address   = {New York},
  year      = {1991}
}

@book{vassallo1991,
  author    = {Vassallo, C.},
  title     = {Optical Waveguide Concepts},
  series    = {Optical Wave Sciences and Technology},
  volume    = {1},
  publisher = {Elsevier},
  address   = {Amsterdam},
  year      = {1991}
}

@article{leung1994qnm,
  author  = {Leung, P. T. and Liu, S. Y. and Young, K.},
  title   = {Completeness and orthogonality of quasinormal modes in leaky
             optical cavities},
  journal = {Phys. Rev. A},
  volume  = {49},
  number  = {4},
  pages   = {3057--3067},
  year    = {1994},
  doi     = {10.1103/PhysRevA.49.3057}
}

@article{ching1998qnm,
  author  = {Ching, E. S. C. and Leung, P. T. and {Maassen van den Brink}, A.
             and Suen, W. M. and Tong, S. S. and Young, K.},
  title   = {Quasinormal-mode expansion for waves in open systems},
  journal = {Rev. Mod. Phys.},
  volume  = {70},
  number  = {4},
  pages   = {1545--1554},
  year    = {1998},
  doi     = {10.1103/RevModPhys.70.1545}
}

@article{sauvan2013qnm,
  author  = {Sauvan, C. and Hugonin, J. P. and Maksymov, I. S.
             and Lalanne, P.},
  title   = {Theory of the spontaneous optical emission of nanosize
             photonic and plasmon resonators},
  journal = {Phys. Rev. Lett.},
  volume  = {110},
  number  = {23},
  pages   = {237401},
  year    = {2013},
  doi     = {10.1103/PhysRevLett.110.237401}
}

@article{lalanne2018qnm,
  author  = {Lalanne, P. and Yan, W. and Vynck, K. and Sauvan, C.
             and Hugonin, J.-P.},
  title   = {Light interaction with photonic and plasmonic resonances},
  journal = {Laser Photonics Rev.},
  volume  = {12},
  number  = {5},
  pages   = {1700113},
  year    = {2018},
  doi     = {10.1002/lpor.201700113}
}

@article{sauvan2022qnm,
  author  = {Sauvan, C. and Wu, T. and Zarouf, R. and Muljarov, E. A.
             and Lalanne, P.},
  title   = {Normalization, orthogonality, and completeness of quasinormal
             modes of open systems: the case of electromagnetism},
  journal = {Opt. Express},
  volume  = {30},
  number  = {5},
  pages   = {6846--6885},
  year    = {2022},
  doi     = {10.1364/OE.443656}
}

@article{kristensen2020qnm,
  author  = {Kristensen, P. T. and Herrmann, K. and Intravaia, F.
             and Busch, K.},
  title   = {Modeling electromagnetic resonators using quasinormal modes},
  journal = {Adv. Opt. Photonics},
  volume  = {12},
  number  = {3},
  pages   = {612--708},
  year    = {2020}
}

@article{haus1991cmt,
  author  = {Haus, H. A. and Huang, W.},
  title   = {Coupled-mode theory},
  journal = {Proc. IEEE},
  volume  = {79},
  number  = {10},
  pages   = {1505--1518},
  year    = {1991}
}

@article{Ge_2014,
author = {Ge, R.-C. and Kristensen, P. T. and Young, J. F. and Hughes, S.},
title = {Quasinormal mode approach to modelling light-emission and propagation in nanoplasmonics},
journal = {New J. Phys.},
volume = {16},
number = {11},
pages = {113048},
year = {2014}
}

\end{document}